\documentclass[prodmode,acmtoms]{acmsmall}
\usepackage{cite}
\usepackage{subcaption}
\usepackage{hyperref}
\usepackage{listings}
\usepackage{color}
\usepackage{amsmath}								
\usepackage{amssymb}								
\usepackage[verbose]{placeins}
\usepackage{stackengine}
\usepackage[binary-units=true]{siunitx}

\usepackage{tikz}
\usetikzlibrary{%
   calc,
   shapes.geometric, 
   arrows,
    decorations.pathreplacing,%
    decorations.pathmorphing,
    patterns,
    decorations.markings,
    external
}
\tikzset{nomorepostaction/.code=\let\tikz@postactions\pgfutil@empty}
\usetikzlibrary{decorations.markings}

\definecolor{plotGray2}{HTML}{bdbdbd}
\definecolor{plotBlue}{HTML}{2C7FB8}
\definecolor{plotGreen}{HTML}{8FDDCB}
\definecolor{plotYellow}{HTML}{EDF8B1}

\definecolor{plotGray}{rgb}{0.95,0.95,0.95}
\definecolor{dkgreen}{rgb}{0,0.6,0}
\definecolor{gray}{rgb}{0.5,0.5,0.5}
\definecolor{mauve}{rgb}{0.58,0,0.82}
\tikzstyle{decision} = [diamond, draw, fill=white, text width=4.5em, text badly centered,
node distance=3cm, inner sep=0pt,minimum width=3cm, minimum height =
3cm,fill=plotGreen]
\tikzstyle{process} = [rectangle, draw, fill=blue!20, text width=5cm, text
centered, rounded corners, minimum height=3em, minimum width =
5.0cm,fill=plotGray]
\tikzstyle{arrow} = [draw, -latex', solid,line width=0.4mm,fill=black]
\tikzstyle{cloud} = [draw, ellipse,fill=red!20, node distance=3cm, minimum height=2em]

\tikzstyle{io} = [trapezium, trapezium left angle=70, trapezium right angle=110, minimum height=1cm, text centered, draw=black,
fill=yellow!30,fill=plotYellow]

\usepackage{scalerel}[2014/03/10]

\acmVolume{V}
\acmNumber{N}
\acmArticle{A}
\acmYear{2016}
\acmMonth{0}
\title{TTC: A high-performance Compiler for Tensor Transpositions}
\author{Paul Springer \affil{AICES, RWTH Aachen} Jeff R. Hammond \affil{Intel Corporation} Paolo Bientinesi
\affil{AICES, RWTH Aachen}}

\ccsdesc[500]{Mathematics of computing~Mathematical software}
\ccsdesc[500]{Theory of computation~Parallel algorithms}
\ccsdesc[500]{Software and its engineering~Compilers}
\ccsdesc[300]{Computing methodologies~Linear algebra algorithms}

\keywords{domain-specific compiler, multidimensional transpositions,
high-performance computing}
\acmformat{Paul Springer, Jeff R.~Hammond and Paolo Bientinesi. 2016. 
    TTC: A high-performance Compiler for Tensor Transpositions
}
\begin{abstract}
We present TTC, an open-source parallel compiler for
multidimensional tensor transpositions.
In order to generate high-performance C++ code, 
TTC explores a number of optimizations, including
software prefetching, blocking, loop-reordering, and explicit vectorization.
To evaluate the performance of multidimensional transpositions across a range
of possible use-cases, we also release a benchmark covering arbitrary transpositions
of up to six dimensions. Performance results show that the routines generated by
TTC achieve close to peak memory bandwidth
on both the Intel Haswell and the AMD Steamroller architectures,
and yield significant performance gains over modern compilers. 
By implementing a set of pruning heuristics, TTC allows users to limit the
number of potential solutions; this option is especially useful when dealing
with high-dimensional tensors, as the search space might become prohibitively
large. Experiments indicate that when only 100 potential solutions are considered, the
resulting performance is about 99\% of that achieved with exhaustive search.

\end{abstract}

\begin{document}
\begin{bottomstuff}
   Author's address: P. Springer (springer@aices.rwth-aachen.de) and P.
   Bientinesi (pauldj@aices.rwth-aachen.de), AICES, RWTH Aachen University,
Schinkelstr.~2, 52062 Aachen, Germany. 
\setcopyright{acmlicensed}
\doi{http://dx.doi.org/10.1145/0000000.0000000}
\issn{0098-3500}
\end{bottomstuff}

\maketitle

\section{Introduction}
\label{sec:intro}



Tensors 
appear in a wide range of applications, including
electronic structure theory~\cite{Bartlett:2007:RMP:CC}, 
multiresolution analysis~\cite{Harrison:2015:arXiv:MADNESS}, 
quantum many-body theory~\cite{pfeifer2014faster}, 
quantum computing simulation~\cite{markov2008simulating},
machine learning~\cite{NVIDIA:2014:arXiv:CUDNN,abaditensorflow,vasilache2014fast}, 
and
data analysis~\cite{kolda2009tensor}. 
While a range of software tools exist for computations
involving one- and two-dimensional arrays, i.e. vectors and matrices,
the availability of high-performance software tools for tensors is
much more limited.
In part, this is due to the combinatorial explosion of
different operations that need to be supported: There are only
four ways to multiply two matrices, but $\binom{m}{k}\binom{n}{k}$ ways to contract
an $m$- and an $n$-dimensional tensor over $k$ indices.
Furthermore, the different dimensions and data layouts that
are relevant to applications are much larger in the case of tensors,
and these issues lead to memory access patterns that are particularly
difficult to execute efficiently on modern computing platforms.

Efficient computation of tensor operations, particularly contractions,
exposes a tension between generality and mathematical expression on
one hand, and performance and software reuse on the other.
If one implements tensor contractions in a naive way---using perfectly-nested loops---the connection
with the mathematical formulae is obvious, but the performance
will be suboptimal in all nontrivial cases.
High performance can be obtained by using the \textit{Basic Linear Algebra Subprograms} (BLAS) \cite{blas3}, but mapping from
tensors to matrices efficiently is nontrivial~\cite{DiNapoli:2014:AMC:tensors,Li:2015:IIA:2807591.2807671} and optimal strategies are unlikely
to be performance portable, due to the ways that multidimensional
array striding taxes the memory hierarchy of modern microprocessors.

A well-known approach for optimizing tensor computations is
to use the level-3 BLAS for contracting matrices at high efficiency,
and to always permute tensor objects into a compatible matrix format.
This approach has been used successfully in the NWChem Tensor
Contraction Engine module~\cite{Hirata:2003:JPCA:TCE,KarolBookChapter},
the Cyclops Tensor Framework~\cite{Solomonik:2014:JPDC:CTF}, and
numerous other coupled-cluster codes dating back more than
25 years~\cite{Gauss:1991:JChP:ACES}.
The critical issue for this approach is the existence of a
high-performance tensor permutation, or tensor transpositions.

While tensor contractions appear in a range of scientific domains (e.g.,~climate simulation \cite{drake1995design}
and multidimensional Fourier transforms~\cite{FFTW,pekurovsky2012p3dfft}),
they are perhaps of greatest importance in
quantum chemistry~\cite{baumgartner2005synthesis,Hirata:2003:JPCA:TCE},
where the most expensive widely used methods---coupled-cluster
methods---consist almost entirely of contractions which deal with 4-,
6-, and even 8-dimensional 
tensors. Such computations consume millions of processor-hours of supercomputing
time at facilities around the world, so any improvement in their performance
is of significant value.

To this end, we developed 
\emph{Tensor Transpose Compiler} (TTC),
an open source code generator
for high-performance multidimensional transpositions that also supports
multithreading and vectorization.\footnote{TTC is available at
{\tt \url{www.github.com/HPAC/TTC}}.}
Together with TTC, we provide a transpose benchmark that can be used to compare
different algorithms and implementations
on a range of multidimensional transpositions.

Let $A_{i_1,i_2,...,i_N}$  be an $N$-dimensional tensor, and
let $\Pi(i_1,i_2,...,i_N)$ denote an arbitrary permutation of the
indices $i_1, i_2, ..., i_N$. The transposition of $A$ into
$B_{\Pi(i_1,i_2,...,i_N)}$ is expressed as 
$B_{\Pi(i_1,i_2,...,i_N)} \gets \alpha \times A_{i_1,i_2,...,i_N}$.
To make TTC flexible and applicable to a
wide range of applications, we designed it to support the class of transpositions
\begin{eqnarray}
   \label{eq:transpose}
   B_{\Pi(i_1,i_2,...,i_N)} \gets \alpha \times A_{i_1,i_2,...,i_N} + \beta \times B_{\Pi(i_1,i_2,...,i_N)},
\end{eqnarray}
where $\alpha$ and $\beta \in \mathbb{R}$; that is, 
the output tensor $B$ can be updated (as opposed to overwritten), and 
both $A$ and $B$ can be scaled.\footnote{An alternative representation for
  Eqn.~\ref{eq:transpose} is
  $B_{i_1,i_2,...,i_N} \gets \alpha \times A_{\widetilde{\Pi}(i_1,i_2,...,i_N)} + \beta \times B_{i_1,i_2,...,i_N}$, 
  where $\widetilde{\Pi}$ is a suitable permutation.} 
As an example, let $A_{i_1,i_2}$ be a two-dimensional tensor, 
$\Pi(i_1,i_2) = (i_2,i_1)$, and $\alpha = 1, \beta = 0$; then Eqn.~\ref{eq:transpose} reduces to 
an ordinary out-of-place 2D matrix transposition of the form $B_{i_2,i_1} \gets
A_{i_1,i_2}$. 

Throughout this
article, we adopt a Fortran storage scheme, that is, the tensors
are assumed to be stored following the indices from left to right (i.e., the leftmost index has stride 1).
Additionally, we use the notation $\pi(i_a) = i_b$ to denote
that the $a$\textsuperscript{th} index of the input operand will become the
$b$\textsuperscript{th} index of the output operand.

For each input transposition, TCC explores
a number of optimizations (e.g.,~blocking, vectorization, parallelization, prefetching,
loop-reordering, non-temporal stores), each of which exposes one or more parameters; to
achieve high-performance, these parameters have
be tuned for the specific hardware on which the transposition will be executed.
Since the effects of such parameters are non-independent, the optimal configuration 
is a point (or a region) of an often large parameter space; indeed, 
as the size of the space grows as $N!$, an exhaustive search is feasible only
for tensors of low dimensionality.
In general, it is widely believed that the parameter space is too complex to design a
\textit{perfect} transpose from first
principles \cite{mccalpin1995automatic,lu2006combining}. 
For all these reasons, 
whenever an exhaustive search is not applicable, 
TTC's generation relies on heuristics.

\begin{figure}[t]
   \centering
   \begin{subfigure}[b]{0.45\textwidth}
      \includegraphics[width=\textwidth]{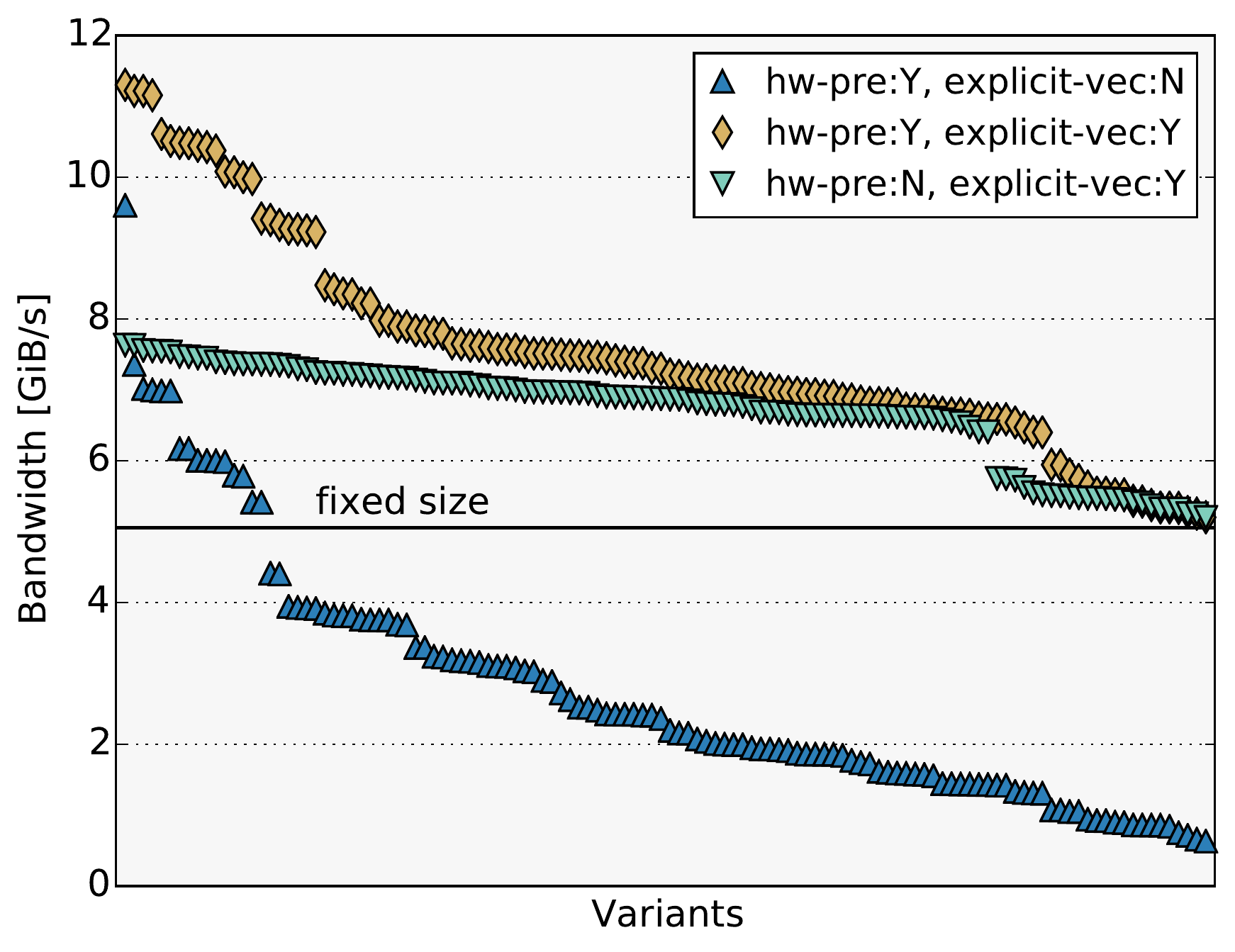}
      \caption{\scriptsize $\Pi(i_1,i_2,i_3,i_4,i_5) = (i_2,i_5,i_3,i_1,i_4)$ } 
      \label{fig:autotuning_a}
   \end{subfigure}
   \begin{subfigure}[b]{0.45\textwidth}
      \includegraphics[width=\textwidth]{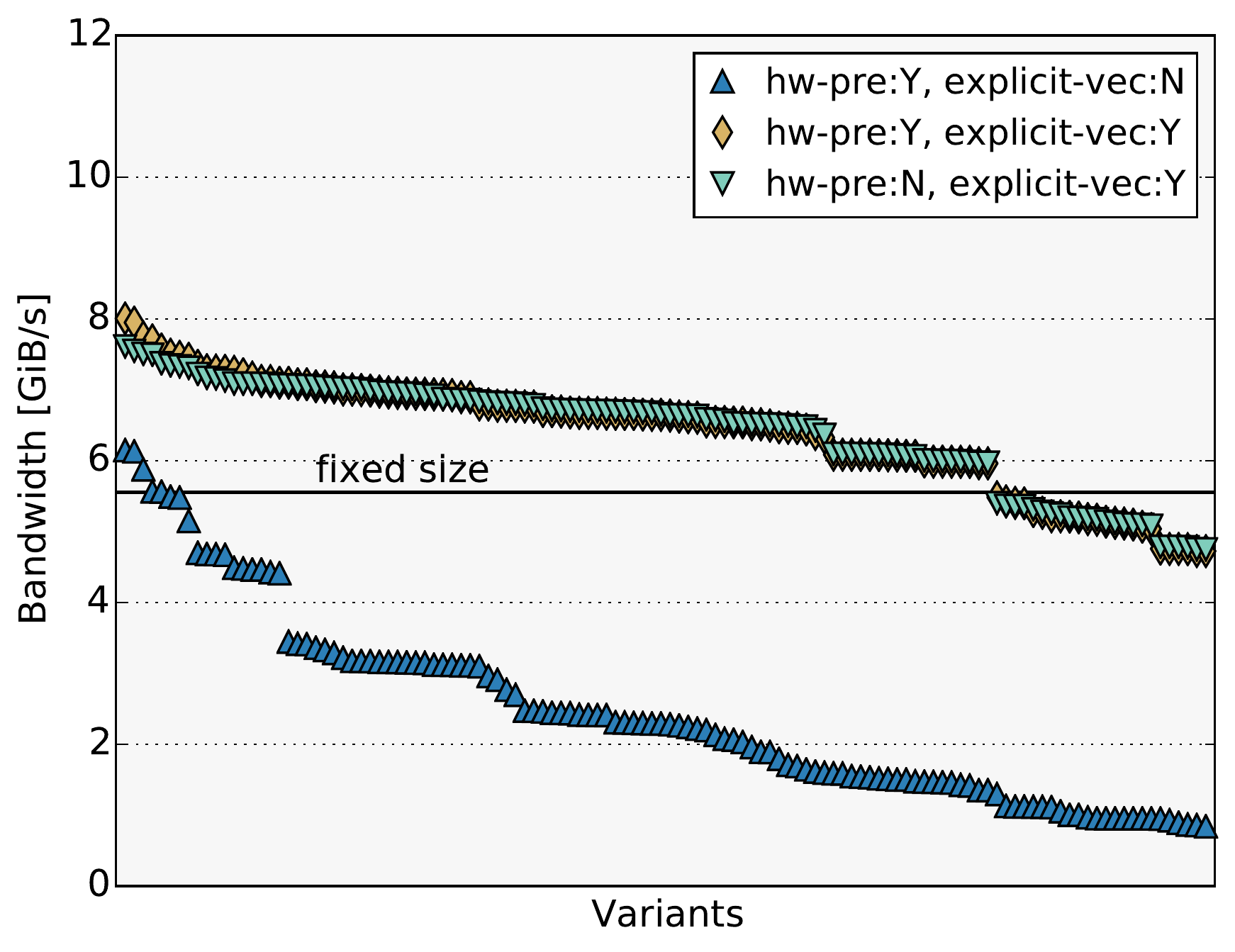}
      \caption{\scriptsize $\Pi(i_1,i_2,i_3,i_4,i_5) = (i_5,i_3,i_1,i_4,i_2)$} 
      \label{fig:autotuning_b}
   \end{subfigure}
   \caption{Single-threaded bandwidth for all the 120 possible loop orders
     of two different 5D transpositions. \textit{explicit-vec} and
     \textit{hw-pre} respectively denote whether explicit
     vectorization and hardware prefetching are enabled (Y) or disabled (N).}
   \label{fig:autotuning}
\end{figure}

In Fig.~\ref{fig:autotuning},
we use two exemplary
5D transpositions (involving tensors of equal volume), as compelling evidence in favor of a
search-based approach.\footnote{The experiments were performed on an Intel Xeon E5-2670 v2 CPU, using
one thread.}
The figures show the bandwidth
attained by each of the
possible $5! = 120$ loop orders\footnote{A $d$ dimensional transposition can be implemented as $d$
perfectly nested loops around an update statement. These loops can be ordered in $d!$
different ways.}---sorted in descending order, from left to right---with and without hardware prefetching
(\textit{hw-pre:N}, \textit{hw-pre:Y}), 
and with and without explicit 
vectorization (\textit{explicit-vec:Y}, \textit{explicit-vec:N}).\footnote{We refer to
\textit{explicit} vectorization to denote 
code which is written with AVX intrinsics; the vectorized versions 
use a blocking of $8\times 8$ (see Section \ref{sec:vectorization}).} 
The compiler-vectorized
code consists of perfectly nested loops including \texttt{\#pragma ivdep}, to assist the
compiler. The horizontal line labeled ``\textit{fixed size}'' denotes the
performance of the compiler-vectorized versions (with hardware prefetching enabled), 
where the size of the tensor was known at compile time; this enabled the
compiler to reorder the loops.

Several observations can be made.
(1) Despite the fact that the two transpositions move exactly the same
amount of data, the resulting top bandwidth is clearly different.
(2) The difference between the leftmost and the rightmost 
datapoints---of any color---provides clear evidence that the loop order has a huge impact on
performance: $9.2\times$ and $4.3\times$ in Fig.~\ref{fig:autotuning_a} and Fig.~\ref{fig:autotuning_b}, respectively. This is in line with the findings of Jeff
Hammond \cite{hammond2009automatically}, who pointed out that the best loop order for a
multidimensional transpose can have a huge impact on performance.
(3) By comparing the leftmost data point of the beige (\includegraphics[height=0.8\baselineskip]{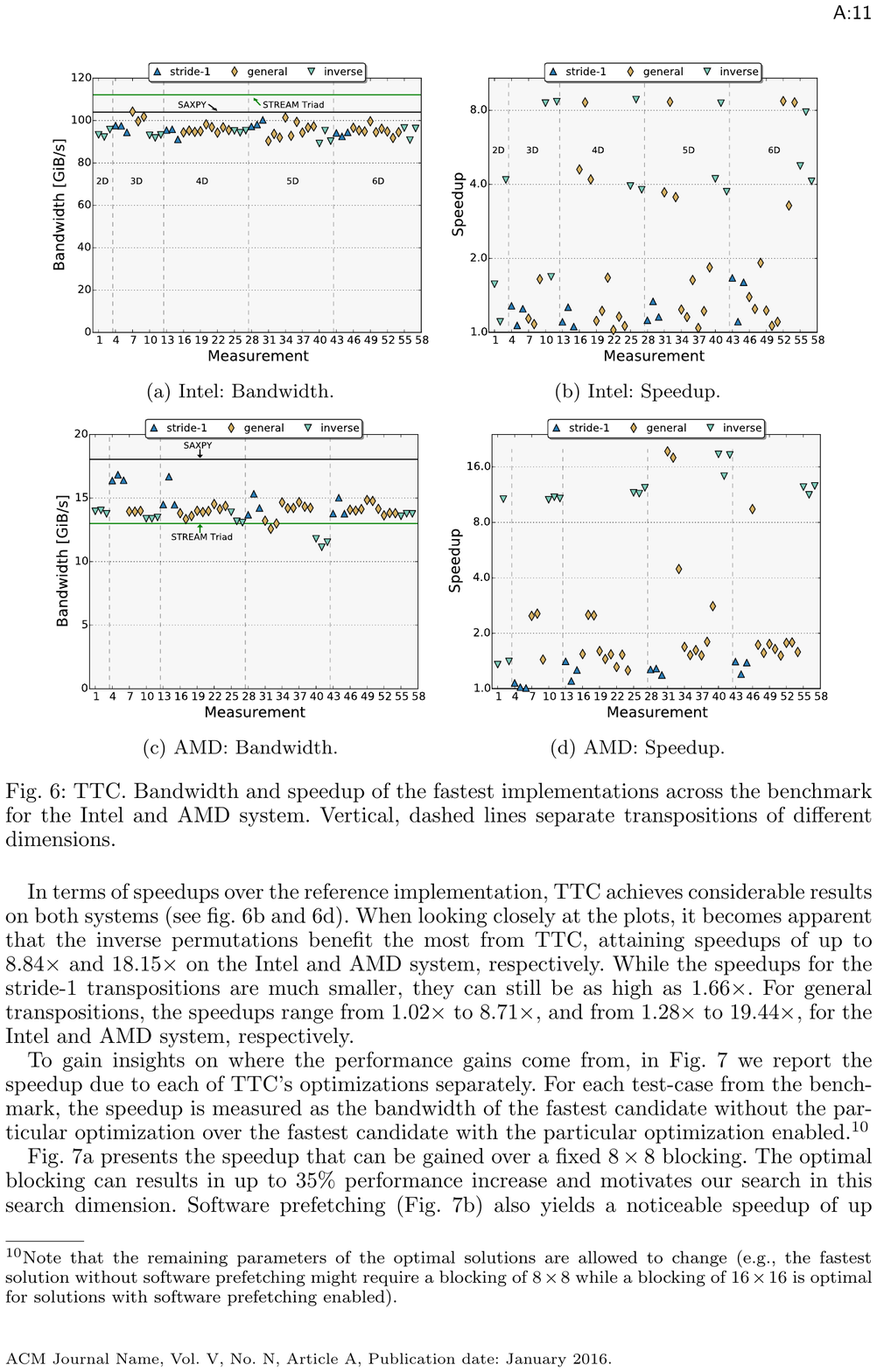}) and blue lines (\includegraphics[height=0.65\baselineskip]{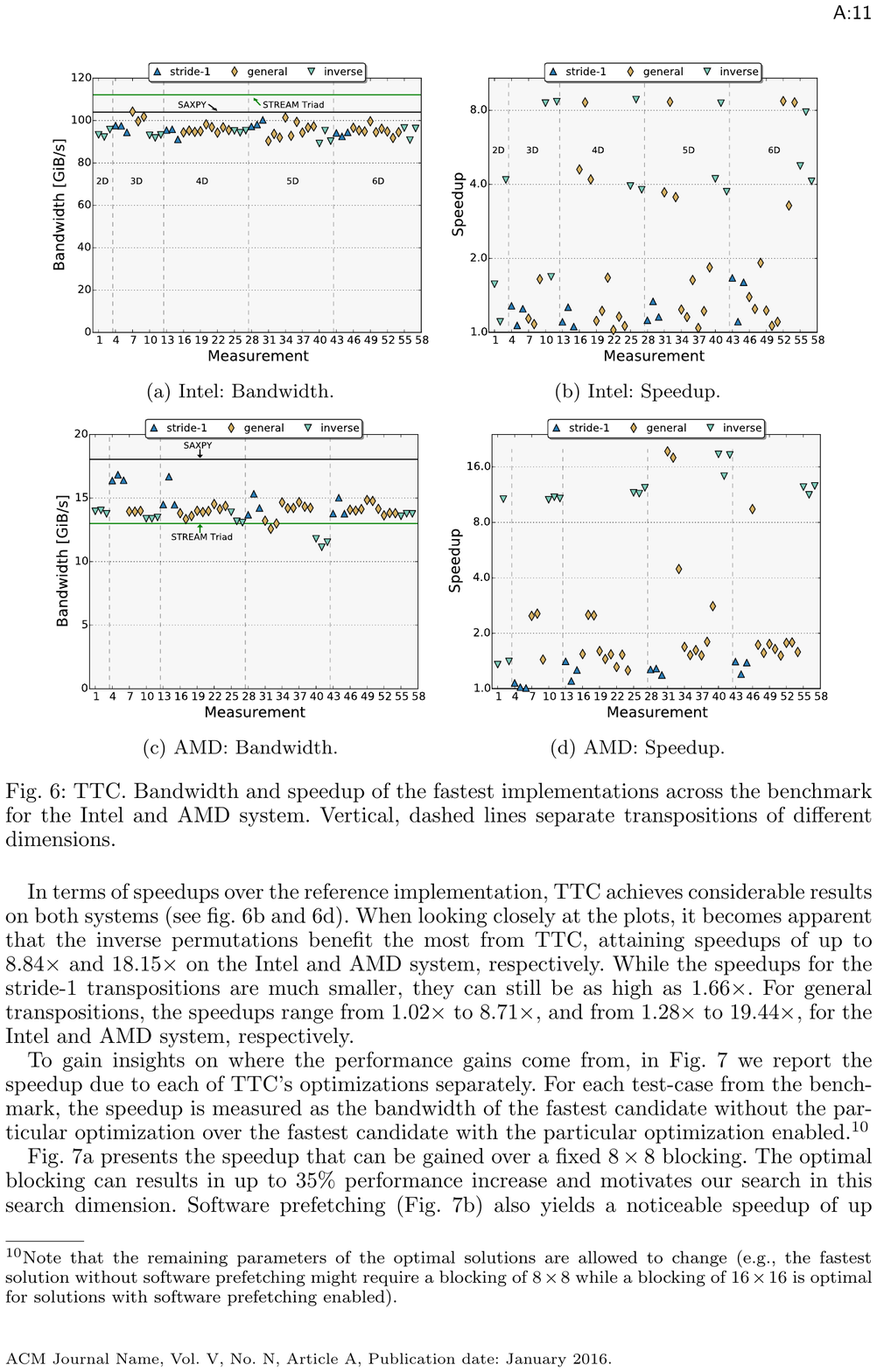}), 
one concludes that the explicit vectorization 
improves the performance 
over the fastest compiler-vectorized version by at least $20\%$.
(4) Since the cyan (\includegraphics[height=0.65\baselineskip]{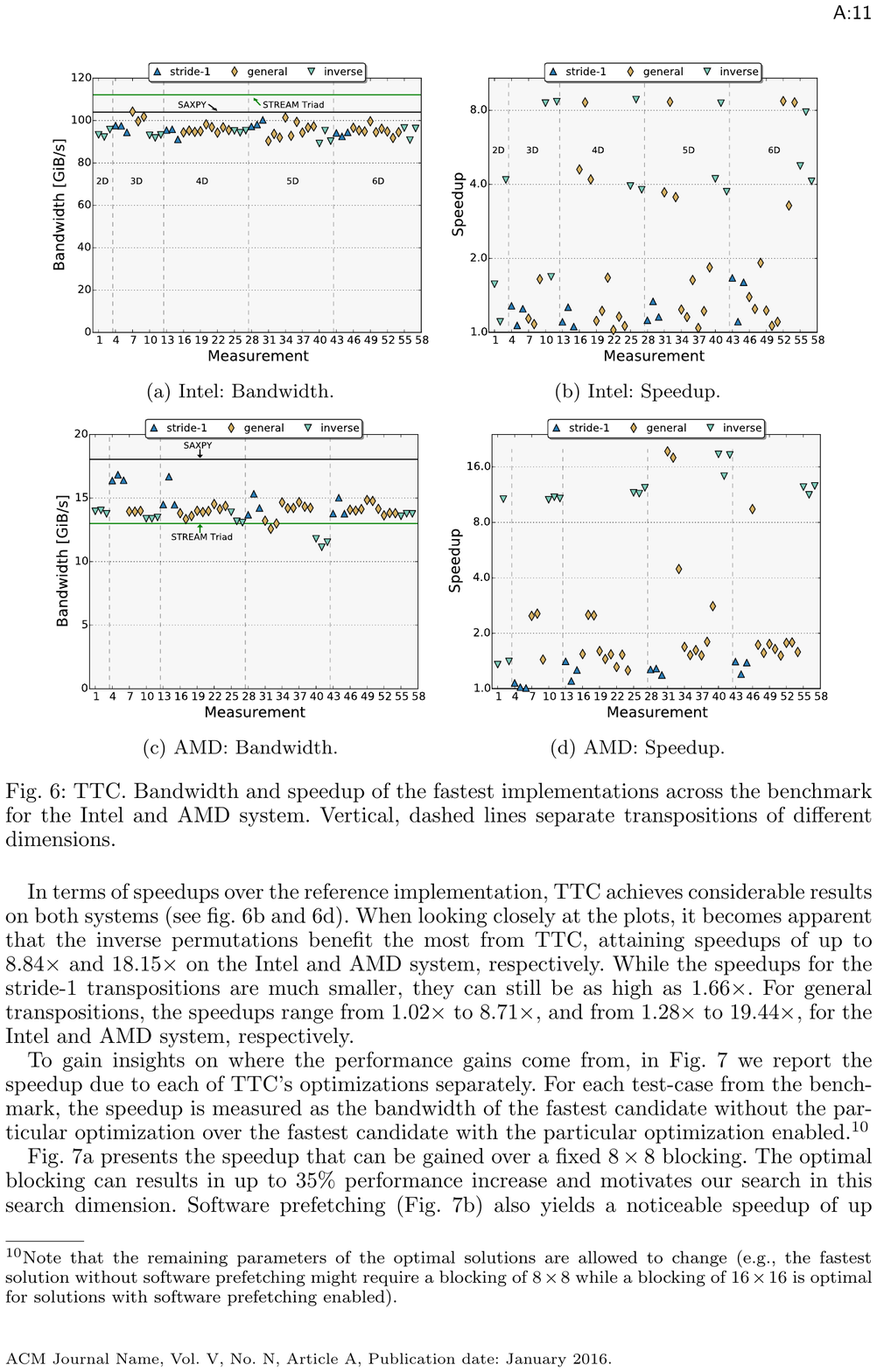}) lines in Fig.~\ref{fig:autotuning_a}
and~\ref{fig:autotuning_b} are practically the same, 
one can conclude that 
the difference in performance between the two
transpositions is due to hardware prefetching.
(5) The difference between the blue line (\includegraphics[height=0.65\baselineskip]{./plots/triangle-up.pdf}) and the horizontal black line 
in Fig.~\ref{fig:autotuning_a} indicates
that when it is possible for the compiler to reorder the loops, the code
generated is much better than most loop orders, but still about 40\% away
from the best one; similarly for Fig.~\ref{fig:autotuning_b}, the
compiler fails to identify the best loop order. 

While observation (5) suggests that modern compilers struggle to find the best
loop order at compile time, an even bigger incentive to adopt a search-based
approach is provided by observation (4), since detailed information about the
mechanism of hardware prefetchers is not well-documented. Moreover, the
implementations of hardware prefetchers can vary between architectures and
manufactures. Hence, designing a generic analytical model for different
architectures seems infeasible at
this point. 

The contributions of this paper can be summarized as follows.
\begin{itemize}
\item 
  We introduce TTC, a high-performance 
  transpose generator
  that supports single, double, single-complex, double-complex and mixed precision data types,
  generates multithreaded C++ code with explicit vectorization for AVX-enabled
  processors, and exploits spatial locality. 
\item 
  We also introduce a comprehensive multidimensional transpose benchmark, 
  to provide the means of comparing different implementations. By means of
  this benchmark, we perform a thorough performance comparison on two architectures.
\item 
  We analyze both the effects of four different optimizations in isolation,
  and quantify the impact of a dramatic reduction of the search space.
\end{itemize}

The remainder of the paper is structured as follows. Section \ref{sec:related} summarizes
the related work on tensor transpositions. In Section \ref{sec:ttc} we introduce
TTC with a special focus on its optimizations. In Section \ref{sec:perf} we present
a thorough performance analysis of TTC's generated implementations---showing both the
attained peak bandwidth as well as the speedups over a realistic baseline implementation.
Finally, Section \ref{sec:conclusion} concludes our findings and outlines future work.

\section{Related Work}
\label{sec:related}

McCalpin et al.~\cite{mccalpin1995automatic} realized that search is necessary for
high-performance 2D transpositions as early as 1995. Their code-generator explored the
optimization space in an exhaustive fashion.

Mateescu et al.~\cite{mateescu2012optimizing} developed a cache model for IBM's Power7 
processor. Their optimizations include blocking, prefetching and data alignment to avoid
conflict-misses. They also illustrate the effect of large TLB\footnote{The translation lookaside buffer, or TLB,
serves as a cache for the expensive virtual-to-physical memory
address translation required to convert software memory addresses
to hardware memory addresses.} page sizes on performance.

Lu et al.~\cite{lu2006combining} developed a code-generator for 2D transpositions using 
both an analytical model and search. They carried out an extensive work covering vectorization,
blocking for both L1 cache and TLB, while parallelization was not explored.

Andre Vladimirov's \cite{vladimirov2013multithreaded}
presented his research on in-place, square transpositions on Intel Xeon and Intel Xeon Phi
processors.

Chatterjee et al.~\cite{Chatterjee:2000ui} investigated the effect of cache and TLB misses
on performance for square, in-place, 2D transpositions. Among other things, they concluded
that ``the limited number of TLB entries can easily lead to
thrashing'' and that ``hierarchical non-linear layouts are
inherently superior to the standard canonical layouts''.

While there has been a lot of research targeted on 2D transpositions
\cite{mccalpin1995automatic,mateescu2012optimizing,goldbogen1981prim,vladimirov2013multithreaded,Chatterjee:2000ui,lu2006combining}
and 3D transpositions \cite{jodra2015efficient,ding2001optimal,van1991fast}, 
higher dimensional transpositions have not yet experienced the same level of
attention. 

Ding et al.~\cite{ding2001optimal,he2002mpi} present an algorithm dedicated to 
multidimensional in-place transpositions. Their approach is optimal with respect to the
number of bytes moved. However, their results suggest that this approach does not yield
good performance if the position of the stride-1 index changes (i.e.,~$\pi(i_1) \neq i_1$).

The work of Wei et al.~\cite{wei2014autotuning} is probably the most complete study of
multidimensional transpositions so far. Their code-generator, which ``uses
exhaustive global search'', explores blocking, in-cache
buffers to avoid conflict misses \cite{gatlin1999memory}, loop unrolling, software
prefetching and vectorization.
The generated code achieves a significant
percentage of the system's \texttt{memcpy} bandwidth on an Intel Xeon and an IBM Power7 node for
cache-aligned transpositions. However, a parallelization
approach is not described, and different loop orders are not considered.

Lyakh et al.~\cite{lyakh2015efficient} designed a generic multidimensional transpose algorithm
and evaluated it across different architectures (e.g.,~Intel Xeon, Intel Xeon Phi,
AMD and NVIDIA K20X). In contrast to our approach, theirs does not rely on search.
Their results suggest that
on both the Xeon Phi as well as the NVIDIA architectures,
there still is a significant performance gap between their
transposition algorithm and a direct copy.

%

\section{Tensor Transpose Compiler}
\label{sec:ttc}
TTC is a domain-specific compiler 
for tensor transpositions of arbitrary dimension. It 
is written in Python and generates high-performance C++ code\footnote{Changing TTC to
generate C code instead of C++ is trivial.}. For a given 
permutation\footnote{We use the terms \textit{permutation} and
\textit{transposition} interchangeably.} and tensor-size, TTC explores a 
search space of possible implementations. These
implementations differ in some properties which have direct effects on
performance e.g.,~loop order, prefetch distance, blocking; each of such properties will be
discussed in the following sections.
Henceforth, we use the term `candidate' for all these implementations;
similarly, we use the term `solution' to denote the fastest (i.e., best
performing) candidate. 

To reduce the compilation time, TTC allows the user to limit the number of candidates to explore; 
by default, TTC explores up to 200 candidates. 
Unless the user specifically wants to explore the entire search space exhaustively, TTC 
applies heuristics to prune the search space and to identify promising candidates
that are expected to yield high performance. A good heuristic should be generic
enough to be applicable on different architectures, and it should
prune the search space to a degree so that the remaining implementations can be
evaluated exhaustively. 
Once the search space has been pruned, TTC generates the C++ routines for all the
remaining candidates, which are compiled and timed.

\begin{table}[hbt]
   \centering
   \begin{tabular}{l | l}
      Argument & Description \\ \hline \hline
      \texttt{--perm=<index1>,<index2>,$\dots$}& 	 permutation$^*$ \\
      \texttt{--size=<size1>,<size2>,$\dots$} & 	 size of each index$^*$\\
      \texttt{--dataType=<s,d,c,z,sd,ds,cz,zc>} & data type of $A$ and $B$ (default:
      \texttt{s})\\
      \texttt{--alpha=<value>} &	 alpha (default: 1.0)\\
      \texttt{--beta=<value>} &	 beta (default: 0.0)\\ 
      \texttt{--lda=<size1>,<size2>,$\dots$}& leading dimension of each index of $A$\\
      \texttt{--ldb=<size1>,<size2>,$\dots$}& leading dimension of each index of $B$\\
      \texttt{--prefetchDistances=<value>,$\dots$} & allowed prefetch distances \\
      \texttt{--no-streaming-stores} & disable non-temporal stores (default: off)\\
      \texttt{--blockings=<H$\times$W>,$\dots$}& block sizes to be explored\\
      \texttt{--maxImplementations=<value>} & max \#implementations (default: 200)
   \end{tabular}
   \caption{TTC's command-line arguments. Required arguments are marked with a $^*$.}
   \label{tbl:arguments}
\end{table} 

The minimum required input to the compiler 
is the actual permutation and the size of each index; 
several additional arguments to pass extra information and to guide the code generation process
can be supplied by command-line arguments;
a subset of the input arguments is listed in Table~\ref{tbl:arguments}.
For instance, by using the \texttt{--lda} and \texttt{--ldb} arguments,
it is possible to transpose tensors that are a portion of larger
tensors. This feature is particularly interesting because it enables TTC to
generate efficient packing routines of scattered data elements into contiguous
buffers; such routines are frequently used in dense linear algebra algorithms
such as a matrix-matrix multiplication \cite{BLIS4}. 
Moreover, TTC supports single-, double-, single-complex- and double-complex data types for
both the input $A$ and the output $B$---via \texttt{--dataType}. TTC is also able to
generate mixed-precision transpositions---where $A$ and $B$ are of different data type (e.g.,
\texttt{--dataType=sd} denotes that $A$ uses single-precision while $B$ uses
double-precision). This feature is again especially interesting in the
context of linear algebra libraries since it allows to implement mixed-precision routines
effortlessly. 
Furthermore, the user can guide the search by
choosing certain blockings, prefetch distances or loop orders---and
thereby reduce the search space. With the argument
\texttt{--maxImplementations}, the user influences the compile time 
by imposing a maximum size to the search space; in the extreme case in which
this flag is set to one, the solution is returned without performing any
search. On the other hand, setting \texttt{--maxImplementations=-1} effectively disables
the heuristics and instructs TTC to explore the search space exhaustively.

\begin{figure}[t]
   \begin{center}
   \tikzsetnextfilename{ttc_overview}
   \hspace*{-0.2cm}
   \begin{tikzpicture}[node distance=2cm,scale=0.7, every node/.style={scale=0.55,font=\large}]
      \node (in) [io, text width=3.0cm] {Permutation, sizes};
      \node (merge) [process, right of=in, xshift=5cm] {(1) Merge indices};
      \node (schedule) [process, below of=merge, yshift=-1.0cm] {(3) Choose
        Loop order};
      \node (add) [process, below of=schedule, yshift=-1.8cm] {(6) Add \mbox{candidate} to queue};
      \node (dec1) [decision, right of=merge,
      xshift=3.8cm,font=\normalsize,scale=1.1] {Solution already known?};
      \node (out) [io, right of=dec1, xshift=4.8cm, text width=3.0cm] {transpose.[cpp/h]};
      \node (blocking) [process, below of=dec1, yshift=-1.0cm] {(2) Choose blocking};
      \node (dec3) [decision, below of=blocking,font=\normalsize,scale=1.1, yshift=-0.9cm] {More combinations?};
      \node (optimizations) [process, below of=in, yshift=-1.0cm] {(4) Choose optimizations};
      \node (dec2) [decision, below of=optimizations,font=\normalsize,scale=1.1, yshift=-0.9cm] {(5) apply heuristics};
      \node (save) [process, below of=out, yshift=-1.0cm] {(9) Store fastest candidate};
      \node (time) [process, below of=save] {(8) Time candidates};
      \node (compile) [process, below of=time] {(7) Compile candidate};

      \draw [arrow] (in) -- (merge);
      \draw [arrow] (merge) -- (dec1);
      \draw [arrow] (dec1) -- node[anchor=south] {Yes} (out);
      \draw [arrow] (dec1) -- node[anchor=east] {No} (blocking);
      \draw [arrow] (blocking) -- (schedule);
      \draw [arrow] (schedule) -- (optimizations);
      \draw [arrow] (optimizations) -- (dec2);
      \draw [arrow] (dec2.east) -- node[anchor=south] {cost okay} ($(add.west)-(0,0.15)$);
      \draw [arrow] ($(add.east)-(0,0.15)$) -- (dec3);
      \draw [arrow] (dec2.south) -- node[anchor=south] {cost too high} (dec3.south);
      \draw [arrow] (dec3) -- node[anchor=east] {Yes} (blocking);
      \draw [arrow] (dec3) -- node[anchor=south] {No} (compile);
      \draw [arrow] (compile) -- (time);
      \draw [arrow] (time) -- (save);
      \draw [arrow] (save) -- (out);
   \end{tikzpicture}
   \caption{Schematic overview of our \textit{tensor transpose compiler} TTC;
     vectorization and parallelization are always enabled.}
   \label{fig:overview}
   \end{center}
\end{figure}
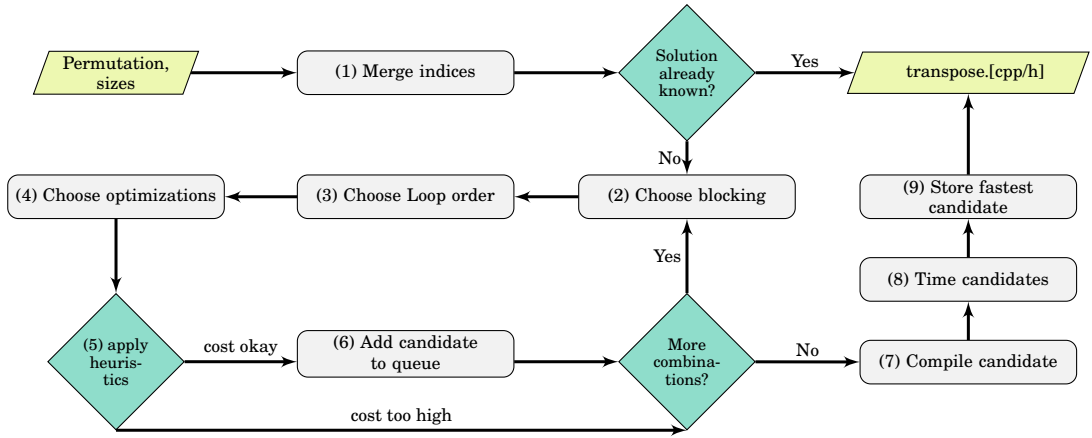

A flowchart outlining the stages (1) - (9) of TTC is shown in Fig.~\ref{fig:overview}.
(1) To reduce complexity, 
TTC starts off by merging indices, whenever possible, in the input and output tensor.
For instance, given the permutation $\Pi(i_1,i_2,i_3)
= (i_2,i_3,i_1)$, the indices $i_2$ and $i_3$ are merged into a new `super index'
$\tilde{i_2}:= (i_2,i_3)$ of the same size as the combined indices (i.e.,~size($\tilde{i_2}$) $=$ size($i_2$)$\times$size($i_3$)); as a consequence, the
permutation becomes $\Pi(i_1,\tilde{i_2}) = (\tilde{i_2},i_1)$.\footnote{Notice that
merging of two indices $i_m$ and $i_{m+1}$ is only possible if $ld(i_{m+1}) =
size(i_{m})ld(i_{m})$ holds, with $size(i)$ and $ld(i)$ respectively denoting the size
and leading-dimension of a given index $i$.}
Next, TTC queries a local SQL database of known/previous solutions, to check whether a
solution for the input transposition already exists; if so, no generation takes place, and
the previous solution is returned. Otherwise, the code-generation proceeds as follows:
(2) one of the possible blockings is chosen, (3) a loop order is selected, and (4) other
optimizations (e.g., software prefetching, streaming-stores) are set. The combination of
the chosen blocking, the loop order and the optimizations uniquely identify a candidate.
After these steps, (5) an estimated cost for the current candidate is calculated. This
cost is used to determine whether the current candidate should be added to a queue of
candidates or if it should be neglected. The aforementioned input argument
\texttt{--maxImplementations} determines the capacity of this queue.

The loop starting and ending at (2) is repeated, 
if different combinations of blockings, loop orders and optimizations are still
possible. 
Once all candidates have been generated, they are
(7) compiled by an external C++ compiler, and (8) timed. 
Finally, (9) the best candidate (i.e.,~the solution) is selected and stored to a \textit{.cpp/h} file
and its timing information as well as its properties (e.g., blocking,
loop order) are saved in the SQL database for future references.

For each and every transposition, TTC explores the following
tuning opportunities:
\begin{itemize}
   \item Explicit vectorization (Section \ref{sec:vectorization})
   \item Blocking (Section \ref{sec:blocking})
   \item Loop-reordering (Section \ref{sec:loop-order})
   \item Software prefetching (Section \ref{sec:prefetch})
   \item Parallelization (Section \ref{sec:parallel})
   \item Non-temporal stores, if applicable
\end{itemize}
The following sections discuss these optimizations in greater detail.
For the remainder of this article, let $w$ denote the vector-width, in elements, for any
given precision and architecture (e.g., $w=8$ for single-precision calculations on an
AVX-enabled architecture).

\subsection{Vectorization}
\label{sec:vectorization}
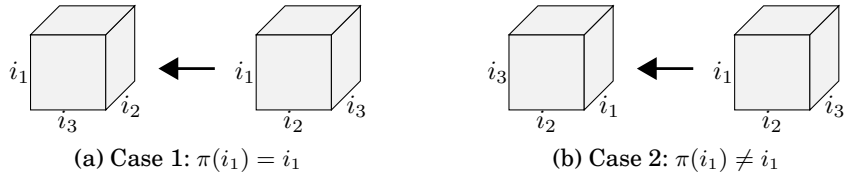
\begin{figure}[t]
   \begin{subfigure}[b]{0.45\textwidth}
   \centering
   \begin{tikzpicture}
\pgfmathsetmacro{\offsetX}{3}
\pgfmathsetmacro{\cubex}{1}
\pgfmathsetmacro{\cubey}{1}
\pgfmathsetmacro{\cubez}{1}
\draw[fill=plotGray] (0,0,0) -- ++(-\cubex,0,0) -- node[left,inner sep = 0] {$i_1$} ++(0,-\cubey,0) -- node[below,inner sep = 0] {$i_3$}++(\cubex,0,0) -- cycle;
\draw[fill=plotGray] (0,0,0) -- ++(0,0,-\cubez) -- ++(0,-\cubey,0) -- node[anchor=north west,inner sep = 0] {$i_2$} ++(0,0,\cubez) -- cycle;
\draw[fill=plotGray] (0,0,0) -- ++(-\cubex,0,0) -- ++(0,0,-\cubez) -- ++(\cubex,0,0) -- cycle;

\draw[->,-triangle 60,line width=1.05pt] (1.45,-0.45,0) to (0.7,-0.45,0);

\draw[fill=plotGray] (\offsetX,0,0) -- ++(-\cubex,0,0) -- node[left,inner sep = 0] {$i_1$} ++(0,-\cubey,0) -- node[below,inner sep = 0] {$i_2$} ++(\cubex,0,0) -- cycle;
\draw[fill=plotGray] (\offsetX,0,0) -- ++(0,0,-\cubez) -- ++(0,-\cubey,0) -- node[anchor=north west,inner sep = 0] {$i_3$} ++(0,0,\cubez) -- cycle;
\draw[fill=plotGray] (\offsetX,0,0) -- ++(-\cubex,0,0) -- ++(0,0,-\cubez) -- ++(\cubex,0,0) -- cycle;
\end{tikzpicture}
\caption{Case 1: $\pi(i_1) = i_1$}
\label{fig:case1}
   \end{subfigure}
\begin{subfigure}[b]{0.45\textwidth}
   \centering
   \begin{tikzpicture}
\pgfmathsetmacro{\offsetX}{3}
\pgfmathsetmacro{\cubex}{1}
\pgfmathsetmacro{\cubey}{1}
\pgfmathsetmacro{\cubez}{1}
\draw[fill=plotGray] (0,0,0) -- ++(-\cubex,0,0) -- node[left,inner sep = 0] {$i_3$} ++(0,-\cubey,0) -- node[below,inner sep = 0] {$i_2$}++(\cubex,0,0) -- cycle;
\draw[fill=plotGray] (0,0,0) -- ++(0,0,-\cubez) -- ++(0,-\cubey,0) -- node[anchor=north west,inner sep = 0] {$i_1$} ++(0,0,\cubez) -- cycle;
\draw[fill=plotGray] (0,0,0) -- ++(-\cubex,0,0) -- ++(0,0,-\cubez) -- ++(\cubex,0,0) -- cycle;

\draw[->,-triangle 60,line width=1.05pt] (1.45,-0.45,0) to (0.7,-0.45,0);

\draw[fill=plotGray] (\offsetX,0,0) -- ++(-\cubex,0,0) --  node[left,inner sep = 0] {$i_1$} ++(0,-\cubey,0) -- node[below,inner sep = 0] {$i_2$} ++(\cubex,0,0) -- cycle;
\draw[fill=plotGray] (\offsetX,0,0) -- ++(0,0,-\cubez) -- ++(0,-\cubey,0) -- node[anchor=north west,inner sep = 0] {$i_3$} ++(0,0,\cubez) -- cycle;
\draw[fill=plotGray] (\offsetX,0,0) -- ++(-\cubex,0,0) -- ++(0,0,-\cubez) -- ++(\cubex,0,0) -- cycle;
\end{tikzpicture}
\caption{Case 2: $\pi(i_1) \neq i_1$}
\label{fig:case2}
   \end{subfigure}
   \caption{Two exemplary 3D transpositions.}
\end{figure}

With respect to vectorization,
we distinguish two different cases:
the stride-1 index (i.e.,~the leftmost index) of the
input and output tensors is constant (see Fig.~\ref{fig:case1}) or not (see Fig.~\ref{fig:case2}).
To achieve optimal performance, these cases require significantly different implementations. 

\subsubsection{Case 1: $\pi(i_1) = i_1$}
When the stride-1 index does not change, vectorization is straightforward. In this case, the
transposition moves a contiguous chunk of memory (i.e.,~the first column)
of the input/output tensor at once as opposed to a single element. Hence, the
operation is essentially a series of nested loops around a \texttt{memcopy} of the size of
the first column. In terms of the memory access pattern, 
this scenario is especially favorable, because of the available spatial data
locality.
Since the vectorization in this case does not require any in-register
transpositions and merely boils down to a couple of vectorized loads and
stores, we leave the vectorization to the compiler, which in this specific scenario is
expected to yield good performance.

\subsubsection{Case 2: $\pi(i_1) \neq i_1$}
In order to take full advantage of the SIMD capabilities of modern processors,
this second case requires a more sophisticated approach. 
Without loss of generality, let us assume that the index $i_b$, with $i_b \neq
i_1$, will become the stride-1 index in $B$ (e.g., $i_b = i_3$ in
Fig.~\ref{fig:case2}). 
Accesses to $B$ are contiguous in memory for
successive values of $i_b$, while accesses to $A$ are contiguous for successive values of
$i_1$. Full vectorization is achieved by unrolling the $i_1$ and $i_b$ loops by 
multiples of $w$ elements, giving raise to an $w\times w$ transpose.
Henceforth, we refer to
such a $w\times w$ tile as a \textit{micro-tile}. The transposition of a micro-tile
is fully vectorized by using an in-register transposition.\footnote{This in-register
transposition---written in AVX intrinsics---is automatically generated by another code-generator of ours
and will be the topic of a later publication.}
Using this scheme, an arbitrarily dimensional out-of-place tensor transposition
is reduced to a series of independent two-dimensional $w\times w$ transpositions, each of which accesses $w$ many
$w$-wide consecutive elements of both $A$ and $B$.

\subsection{Blocking}
\label{sec:blocking}
\begin{figure}[t]
   \centering
      \includegraphics[height=4cm]{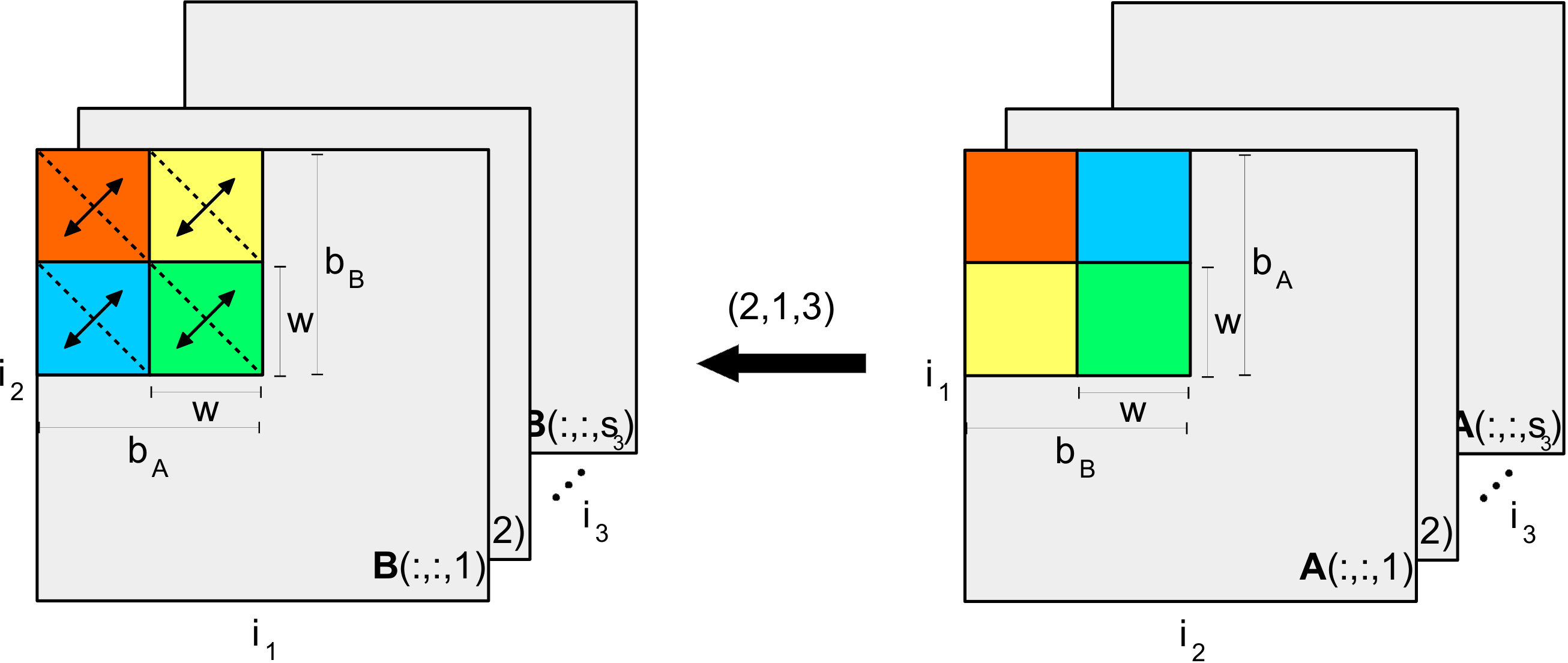}
      \caption{Overview of the blocking mechanism for the 
        permutation $\Pi(i_1,i_2,i_3) = (i_2,i_1,i_3)$, with $b_A = b_B = 2w$.}
      \label{fig:blockingOverview}
\end{figure}

In addition to the $w\times w$ micro-tiles, we introduce a second level of
blocking to further increase locality. The idea is to combine multiple micro-tiles into
a so-called \textit{macro-tile} of size $b_A\times b_B$, where $b_A$ and $b_B$
correspond to the blocking in the stride-1 index of $A$ and $B$, respectively.\footnote{In
the case of $\pi(i_1) = i_1$, such a blocking would not make sense; hence, the
blocking takes place along
the second index of $A$ and $B$. For instance, for $B_{i_1,i_3,i_2}
\gets A_{i_1,i_2,i_3}$, the blocking involves $i_2$ and $i_3$.}
This approach is illustrated in Fig.~\ref{fig:blockingOverview}.

By default, TTC explores the search space of all macro-tiles of size $b_A\times
b_B$ with $b_A, b_B \in \{w,2w,3w,4w\}$. The flexibility of supporting 
multiple sizes of macro-tiles has several desirable advantages: first, it
enables TTC to adapt to different memory systems, which might favor contiguous
writes (e.g., $16\times32$) over contiguous reads (e.g., $32\times16$) and
vice versa; second, it implicitly exploits architectural features such as
\textit{adjacent cacheline prefetching}, and cacheline-size (e.g., $16$
elements for single precision for modern x86 CPUs); finally, it reduces false-sharing
of cache lines between different threads in a parallel setting.

If desired by the user, TTC can effectively prune the search space and only evaluate the
performance of a subset of the available tiles. We designed a heuristic
which ranks the blockings for a given transposition and size. 
Specifically, the blocking is chosen such that (1) $b_A$ and $b_B$ are both
multiples of the cacheline-size (in elements), and that (2) the remainder $r^i = S^i_1
\text{(mod) } b_i, i \in \{A,B\}$, with $S^i_1$ being the size of the stride-1
index of tensor $i \in \{A,B\}$, is minimized. 
The quality of this heuristic is demonstrated in Section \ref{sec:heuristics}.

\subsection{Loop order}
\label{sec:loop-order}

As Fig.~\ref{fig:autotuning} already suggested, the choice of the proper loop order
has a significant influence on performance. 
Since the number of available orderings for a tensor with $d$ dimensions is $d!$,
determining the best loop order is by exhaustive search is expensive even
for modest values of $d$.

Our heuristic to choose the loop order is designed to increase data locality in
both $A$ and $B$.
This strategy fulfills multiple purposes:
(1) it reduces cache- and TLB-misses, and
(2) it reduces the stride within the innermost loop.
The latter is especially important because large strides can prevent modern
hardware prefetchers from learning the memory access patterns.
For instance, the maximal stride supported by hardware prefetchers of
Intel Sandy Bridge CPUs is limited to $\SI{2}{\kibi\byte}$ \cite{intelOpt}.
Other aspects of the hardware implementation affect the cost of
different loop orders; for example, the write-through cache policy
of the IBM Blue Gene/Q architecture makes it extremely important
to exploit write locality, since writing to a cache line
evicts it from cache~\cite{bgqOpt}.
The reader interested in further details on this heuristic
is referred to the available source-code at
{\tt \url{www.github.com/HPAC/TTC}}.

\subsection{Software Prefetching}
\label{sec:prefetch}
Software prefetching is only enabled for the case of $\pi(i_1)\neq i_1$;
indeed, the memory access pattern for $\pi(i_1) = i_1$ 
is so regular that it should be easily caught by the hardware prefetcher.

\begin{figure}[t]
   \centering
      \includegraphics[height=4.0cm]{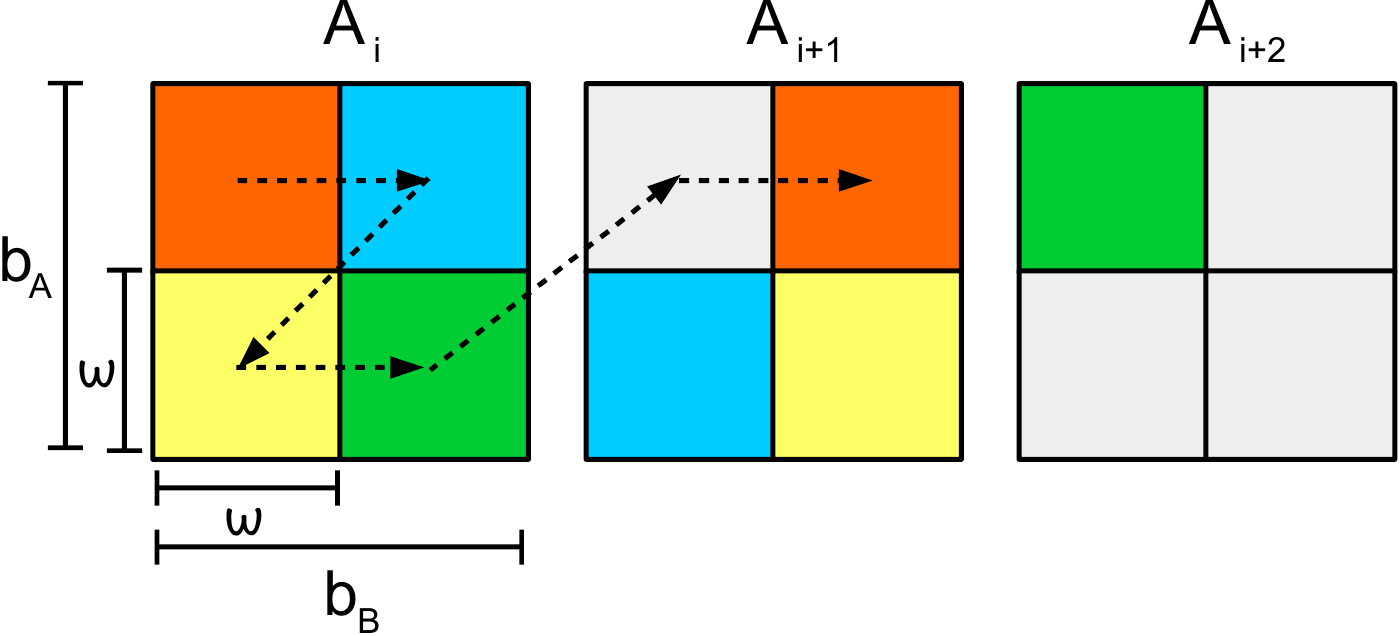}
      \caption{Overview of the software prefetching mechanism for a distance
        $d=5$. Arrows denote the order in which the micro-tiles are being processed.}
      \label{fig:prefetchOverview}
\end{figure}

We designed the software prefetching to operate on micro-tiles; hence, a given prefetch distance $d$ has the same
meaning irrespective of the chosen macro-blocking. The prefetching mechanism is
depicted in Fig.~\ref{fig:prefetchOverview}. TTC always prefetches entire $w\times w$ micro-tiles. Before
transposing the current micro-tile $j$, TTC prefetches the micro-tile which is at distance $d$
ahead of the current tile. This is illustrated by the colors in Fig.~\ref{fig:prefetchOverview}, where the macro-tile contains 
$n = 4$ micro-tiles:
before processing the orange $w\times w$ block of
the current macro-tile $A_i$, TTC already prefetches the orange micro-tile of the
corresponding macro-tile $A_p$, with $p = i +
\lfloor(j+d)/n\rfloor$ (i.e.,~$A_{i+1}$ in Fig.~\ref{fig:prefetchOverview}).

\subsection{Parallelization}
\label{sec:parallel}
\begin{center}
   \begin{minipage}{0.75\textwidth}
      \begin{lstlisting}[caption={Parallel code generated by TTC for the permutation ($i_3,i_2,i_1$).}, label=lst:parallel]
//variable declaration
...

//main loops
#pragma omp parallel
#pragma omp for collapse(3) schedule(static)
   for(int i3 = 0; i3 < size2 - 15; i3+= 16)
      for(int i1 = 0; i1 < size0 - 31; i1+= 32)
         for(int i2 = 0; i2 < size1; i2+= 1)
            transpose32x16(&A[i1 + i2*lda1 + i3*lda2], lda2, 
                          &B[i1*ldb2 + i2*ldb1 + i3], ldb2, alpha, beta);
//remainder loops
...
\end{lstlisting}
\end{minipage}
\end{center}

Listing \ref{lst:parallel} contains the generated parallel code\footnote{We 
do not present the code that handles the remainder generated when
$b_A$ or $b_B$ do not evenly divide the leading dimension of $A$ or $B$.} for a given tensor
transposition with software prefetching disabled. The \texttt{collapse} clause increases the available
parallelism and improves load balancing among the threads. Each thread has to
process roughly the same amount of $b_A\times b_B$ tiles.

A detailed description of the parallelization of the prefetching algorithm is beyond the
scope of this paper. However, the overall idea of prefetching the blocks remains identical to the scalar version (see
previous section). The only difference is that each thread has to keep track of
the tiles it will access in the near future 
in a local data structure; for this task, we use a queue of tiles. The interested reader is again
referred to the source-code for further details.

\section{Performance Evaluation}
\label{sec:perf}
We evaluate the performance of TTC on two 
different systems,
\textit{Intel} and \textit{AMD}.
The \textit{Intel} system 
consists of two \textit{Intel Xeon E5-2680 v3} CPUs (with 12 cores each) based on
the \textit{Haswell} microarchitecture. For all measurements, ECC is enabled,
and both \textit{Intel Speedstep} and \textit{Intel TurboBoost} are disabled.
The compiler of choice 
is the \textit{Intel icpc 15.0.4} with flags \texttt{-O3 -openmp -xhost}. 
Unless otherwise mentioned, this is the default configuration and system for
the experiments.
The \textit{AMD} system consists of a single \textit{AMD A10-7850K} APU with 4 cores based
on the \textit{steamroller} microarchitecture. The compiler for this system is 
\textit{gcc 5.3} with flags \texttt{-O3 -fopenmp -march=native}. All
measurements are based on 24 threads and 4 threads for the Intel and AMD
system, respectively (i.e., one thread per physical core).

Experimental results suggest that optimal performance is attained with one
thread per physical core. We also experimented with thread affinity on both systems.\footnote{On
the Intel systems setting the affinity to
\texttt{KMP\_AFFINITY=granularity=fine,compact,1,0}
(i.e.,~hyper-threading will not be used) yields optimal results. The performance
on the AMD system, on the other hand, was not sensitive to the thread affinity as long
as the threads were not pinned to the same physical core.}

The reported bandwidth $\text{BW}(x)$ for solution $x$ is computed as
\begin{equation}
   \text{BW}(x) = \frac{3 \times S}{2^{30} \times \text{Time}(x)}\text{ } \SI{}{\gibi\byte}/s,
\end{equation}
where $S$ is the size in bytes of the tensor;
the prefactor $3$ is due to the fact
that since in all our measurements $B$ is updated (i.e., $\beta \neq 0$, see
Eqn.~\ref{eq:transpose}), one has to account for reading $B$ as well. 

The reported memory bandwidth by the STREAM benchmark \cite{McCalpin1995} for
the \textit{Intel} (\textit{AMD}) system
is 105.6 (12.2), 105.9 (12.2), 111.6 (13.1) and 112.2 (13.0)
GiB/s for the \textit{copy} ($\mathbf{b} \gets \mathbf{a}$), \textit{scale} ($\mathbf{b} \gets \alpha
\mathbf{a}$),
\textit{add} ($\mathbf{c} \gets \mathbf{a} + \mathbf{b}$) and \textit{triad} ($\mathbf{c} \gets \alpha
\mathbf{a} + \mathbf{b}$)
test-cases, respectively.

\subsection{Transposition Benchmark}
To evaluate the performance of multidimensional transpositions across a range
of possible use-cases, we designed a synthetic benchmark.
The benchmark comprises 19 different transpositions,\footnote{
One 2D transposition, three 3D, and five each for 4D, 5D and 6D.}
chosen so that no indices can be merged.
For each tensor dimension (2D--6D), we included the inverse permutation (e.g., $B_{i_3,i_2,i_1} \gets
A_{i_1,i_2,i_3}$, $B_{i_4,i_3,i_2,i_1} \gets A_{i_1,i_2,i_3,i_4}$, \dots), 
and exactly one permutation for which the stride-1 index
does not change. These two scenarios typically cover both ends of the
spectrum, yielding the worst and the best performance, respectively. 

In the benchmark, each transposition is evaluated in three different
configurations---for a total of 57 test cases: one where all indices are of roughly the same size, and two
where the tensors have a ratio of $6$ between the largest and smallest
index. 
The desired volume of the tensors across all dimensions 
are roughly the same and can be chosen by the user; 
in our experiments, we fixed it to $\SI{200}{\mebi\byte}$, which is bigger than
any L3 cache in use today.

The benchmark is publicly available at {\tt \url{www.github.com/HPAC/TTC/benchmark}}.

\subsection{TTC-generated code}
We now present the performance of the fastest implementations---generated by
TTC---for all transpositions in the benchmark.
Furthermore, we analyze the influence of the individual optimizations
(blocking, loop-reordering, software prefetching, and explicit vectorization)
on the performance.

\begin{figure}[t]
   \centering
   \begin{subfigure}[b]{0.45\textwidth}
   \centering
      \includegraphics[height=5cm]{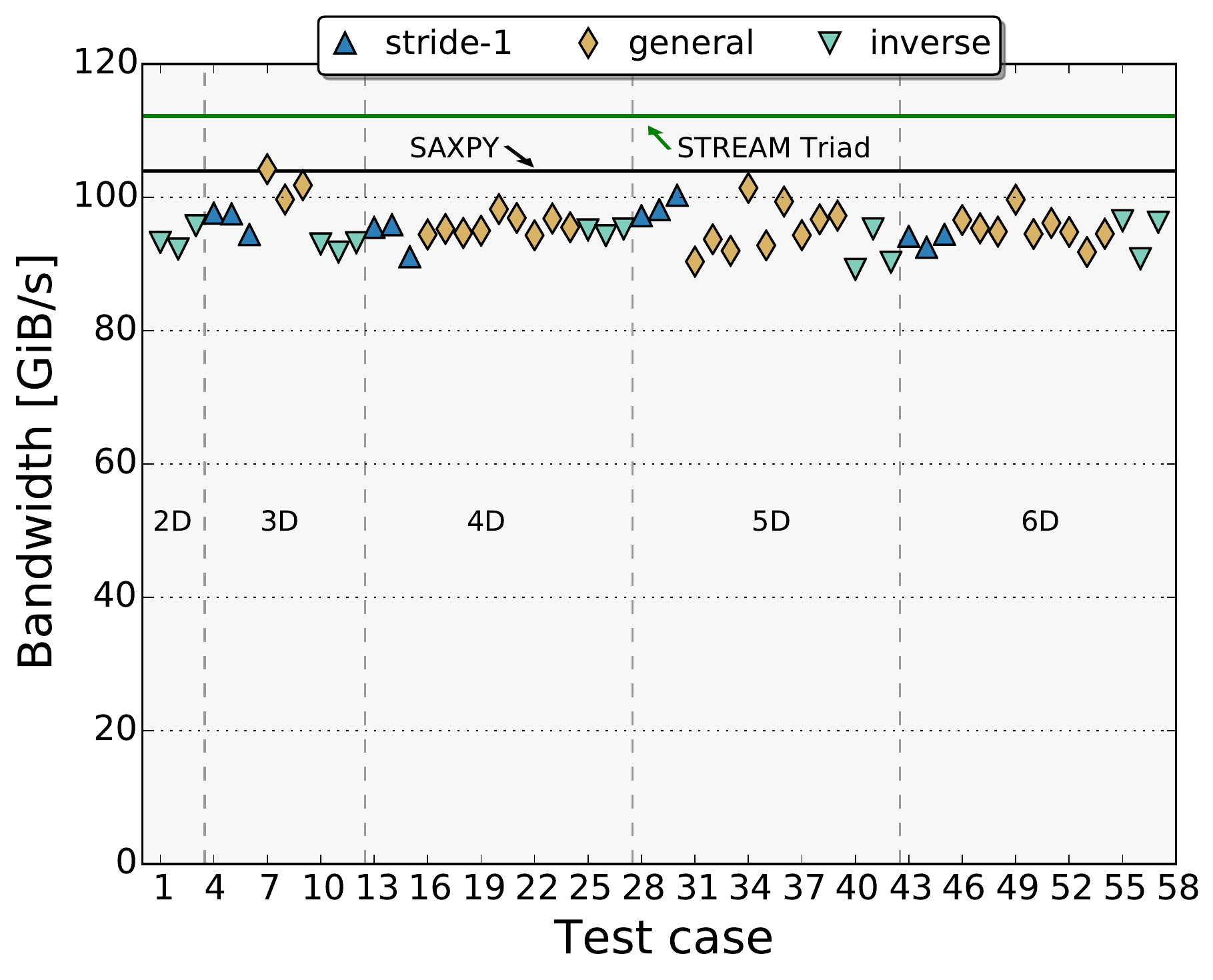}
      \caption{Intel: Bandwidth.}
      \label{fig:benchmarkBandwidthIntel}
   \end{subfigure}
   \begin{subfigure}[b]{0.45\textwidth}
   \centering
      \includegraphics[height=5cm]{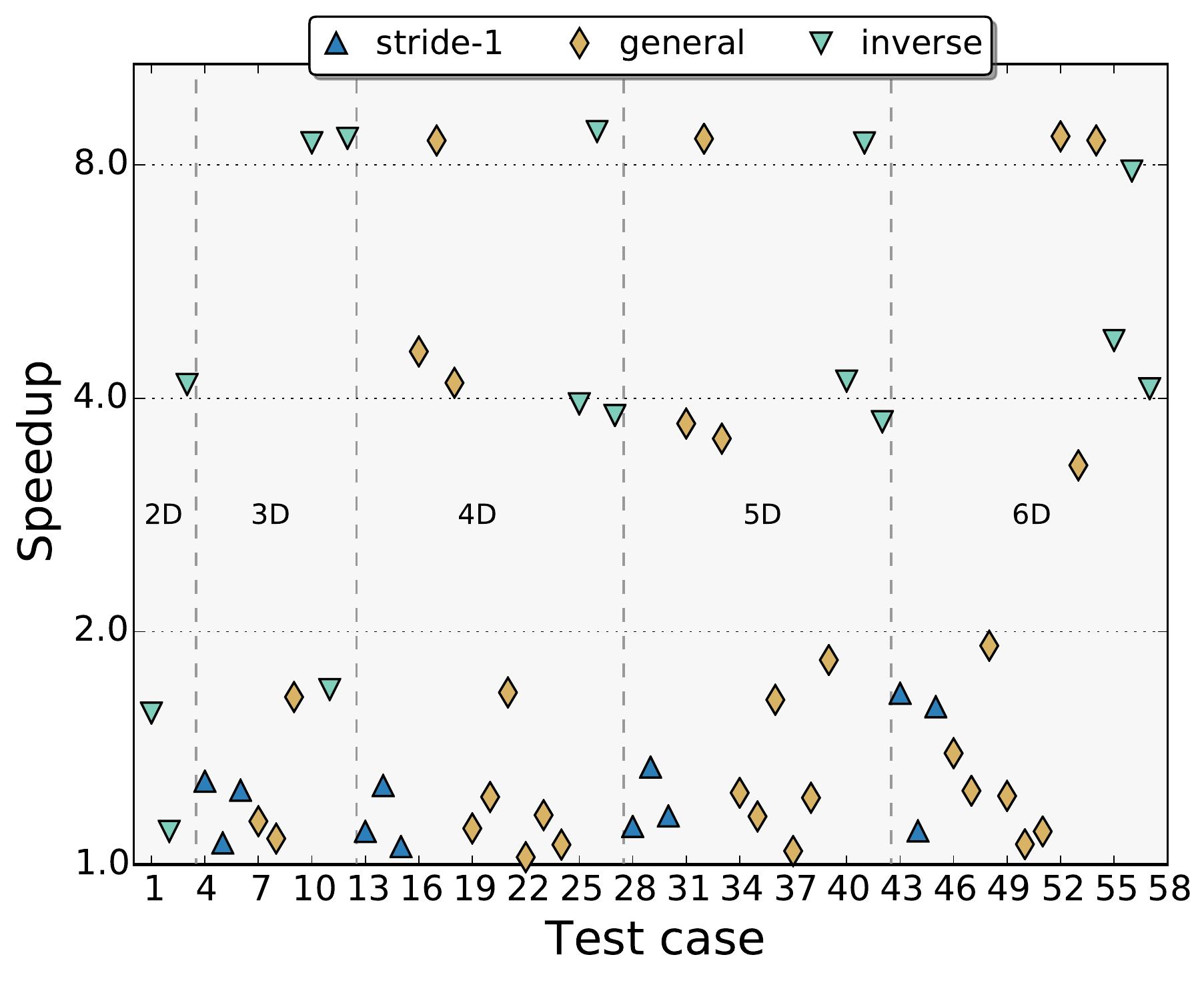}
      \caption{Intel: Speedup.}
      \label{fig:benchmarkSpeedupIntel}
   \end{subfigure}\\
   \begin{subfigure}[b]{0.45\textwidth}
   \centering
      \includegraphics[height=5cm]{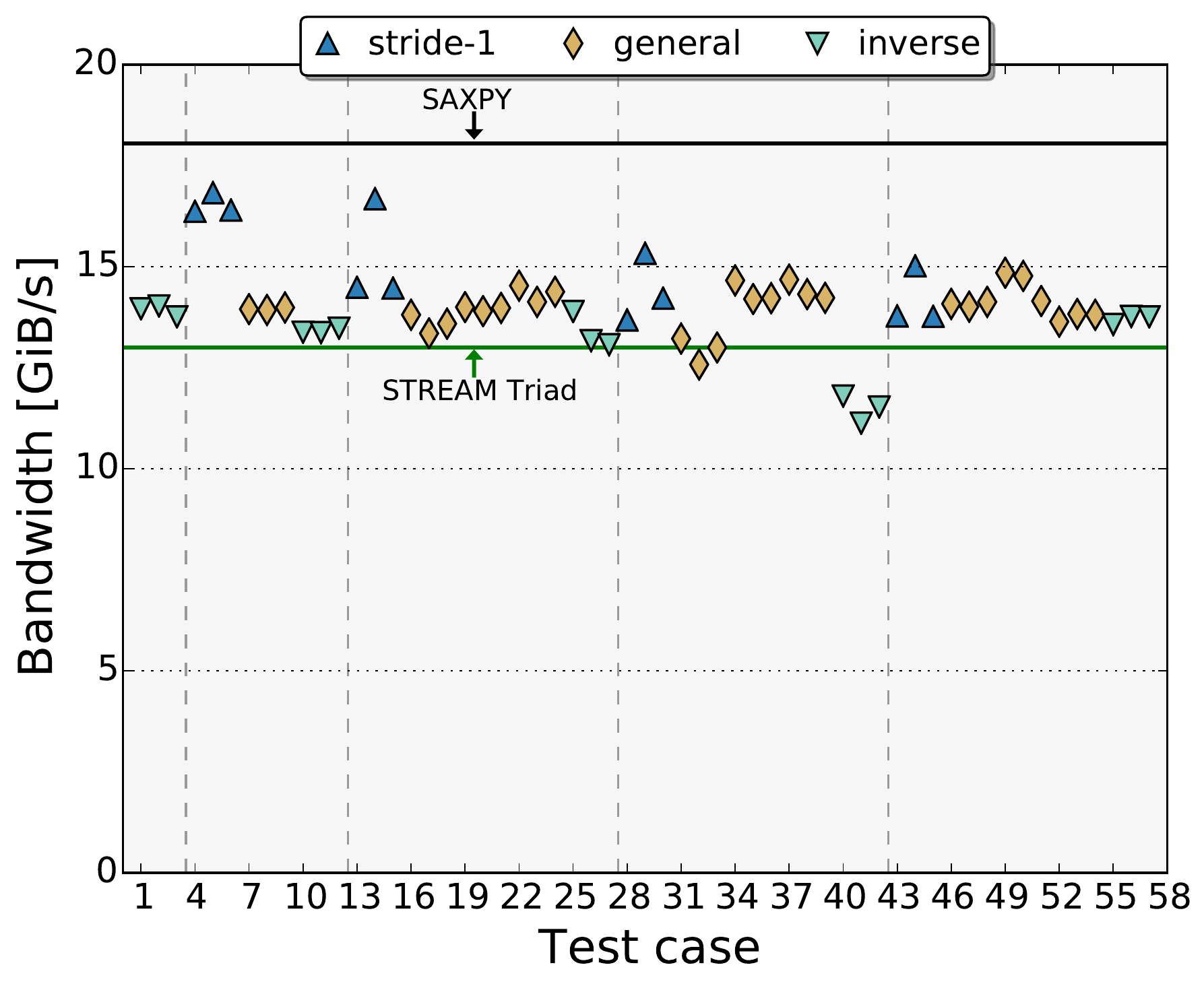}
      \caption{AMD: Bandwidth.}
      \label{fig:benchmarkBandwidthAMD}
   \end{subfigure}
   \begin{subfigure}[b]{0.45\textwidth}
   \centering
   \includegraphics[height=5cm]{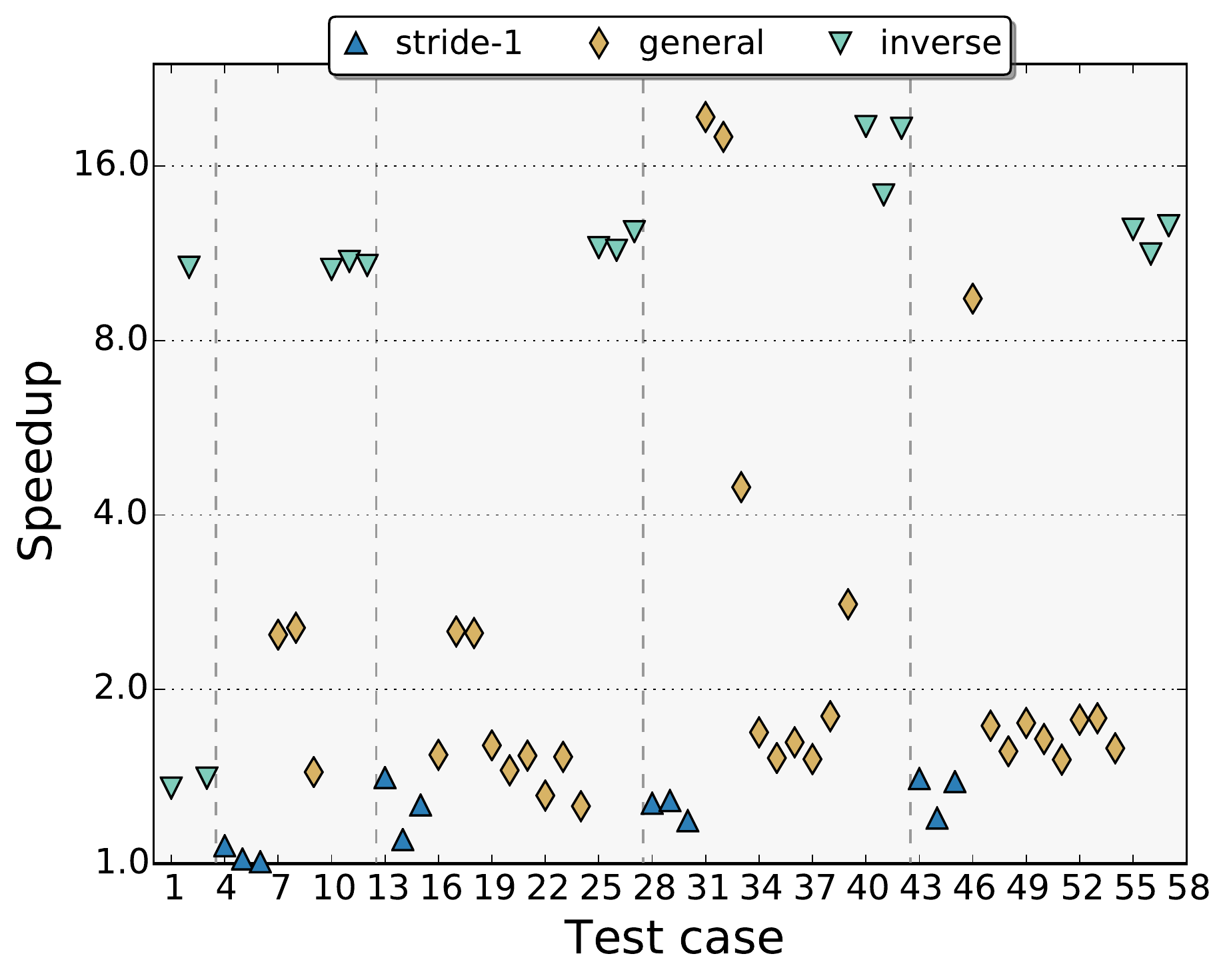}
      \caption{AMD: Speedup.}
      \label{fig:benchmarkSpeedupAMD}
   \end{subfigure}
   \caption{TTC. Bandwidth and speedup of TTC's fastest solution 
     across the benchmark for the Intel and AMD system. The vertical lines identify the dimensionality
      of the tensors.}
   \label{fig:performanceBenchmark}
\end{figure}

Fig.~\ref{fig:performanceBenchmark} illustrates the attained bandwidth and speedup across the
benchmark for the Intel and AMD systems. The speedups are measured over
a reference routine consisting of $N$ perfectly nested loops annotated with
both \texttt{\#pragma omp parallel for collapse($N-1$)} on the outermost loop,
and \texttt{\#pragma omp simd} on the innermost loop.
Moreover, the loop order for the reference version is chosen such that the
output tensor $B$ is accessed in a perfectly linear fashion; this loop order
reduces false sharing of cache lines between the threads. With this setup,
the compiler is assisted as much as possible
to yield a competitive routine.

Figs.~\ref{fig:benchmarkBandwidthIntel} and \ref{fig:benchmarkBandwidthAMD} show
the attained bandwidth of the TTC-generated solutions across the benchmark for
the Intel and AMD system, respectively. The transpositions are classified
in three categories: \textit{stride-1}
(\includegraphics[height=0.65\baselineskip]{./plots/triangle-up.pdf}),
\textit{inverse} (\includegraphics[height=0.65\baselineskip]{./plots/triangle-down.pdf}), and
\textit{general}
(\includegraphics[height=0.8\baselineskip]{./plots/dimond.pdf}), respectively
denoting those permutations in which the first 
index does not change, the inverse permutations, and those transpositions which do
not fall into either of the previous two categories. In addition to the
bandwidth, these figures also report the STREAM-triad
bandwidth (solid green line), as well as the bandwidth of a SAXPY
(i.e.,~single-precision vector-vector addition of the form $\mathbf{y} \gets \alpha
\mathbf{x} + \mathbf{y}$, $\alpha
\in \mathbb{R}, \mathbf{x,y} \in \mathbb{R}^n$, see solid black line). 
The figures
illustrate that TTC achieves a significant fraction of the SAXPY-bandwidth on both architectures
(the average across the entire
benchmark is $91.68\%$ 
and $78.30\%$ for the Intel 
and AMD systems, respectively). 
With the exception of some performance-outliers on the AMD system, the performance of TTC
is stable across the entire benchmark. 
 
It is interesting to note that on the AMD system, 
TTC attains much higher bandwidth than the STREAM-triad benchmark.
This phenomenon is due the fact that the STREAM benchmark does not
account for the write-allocate traffic
.

In terms of speedups over the reference implementation, TTC achieves 
considerable results
on both
systems (see Fig.~\ref{fig:benchmarkSpeedupIntel} and
\ref{fig:benchmarkSpeedupAMD}). When looking closely at the plots, 
it becomes apparent that the inverse permutations benefit the
most from TTC, attaining speedups of up to $8.84\times$ and $18.15\times$ on the
Intel and AMD system, respectively. While the speedups for the
stride-1 transpositions are much smaller,
they can still be as high as $1.66\times$. 
For general transpositions, the speedups 
range from $1.02\times$ to $8.71\times$, and from $1.28\times$ to $19.44\times$, for
the Intel and AMD system, respectively.

\begin{figure}[t]
   \centering
   \begin{subfigure}[b]{0.45\textwidth}
   \centering
      \includegraphics[height=4.5cm]{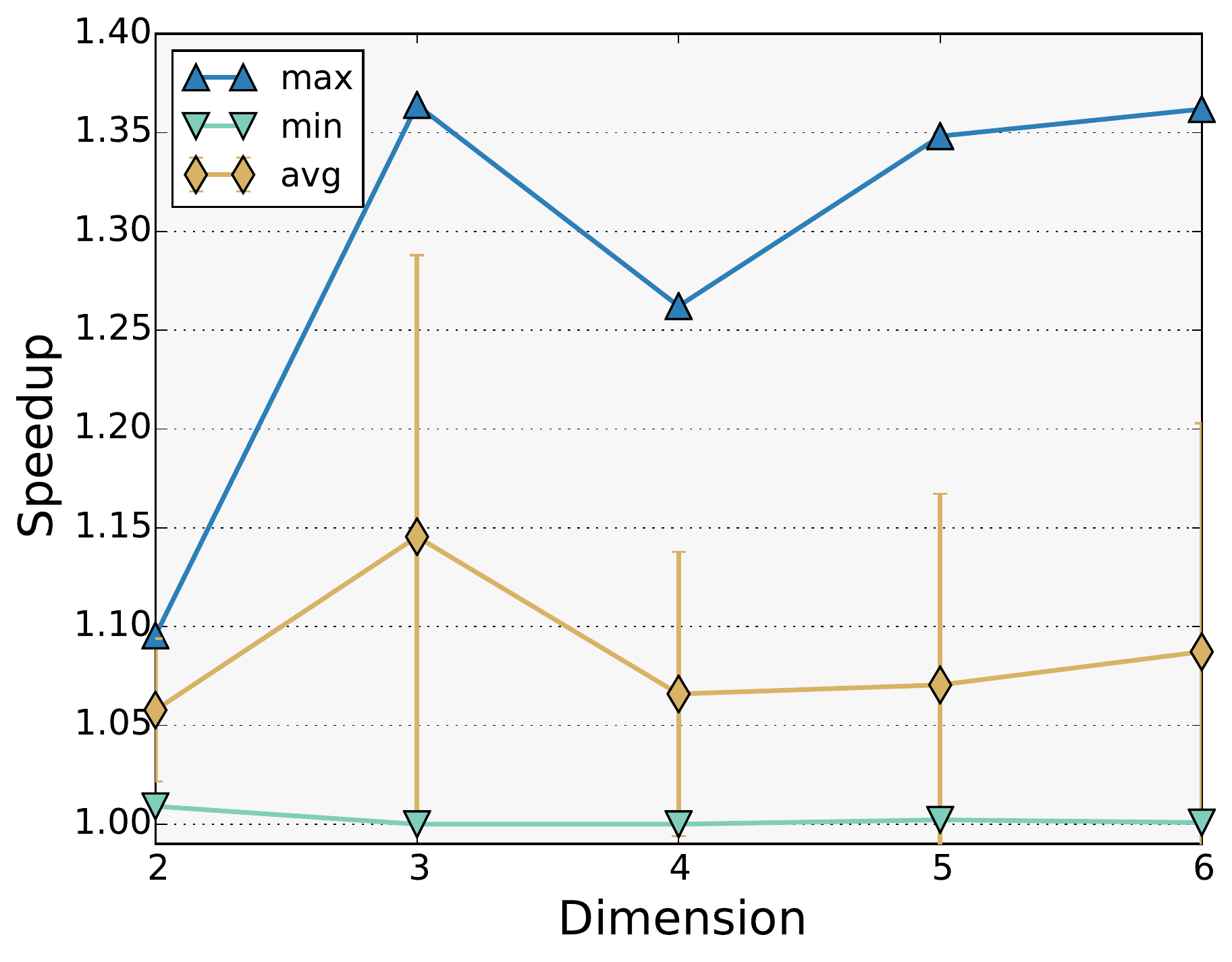}
      \caption{Blocking}
      \label{fig:blockingSpeedup}
   \end{subfigure}
   \begin{subfigure}[b]{0.45\textwidth}
   \centering
      \includegraphics[height=4.5cm]{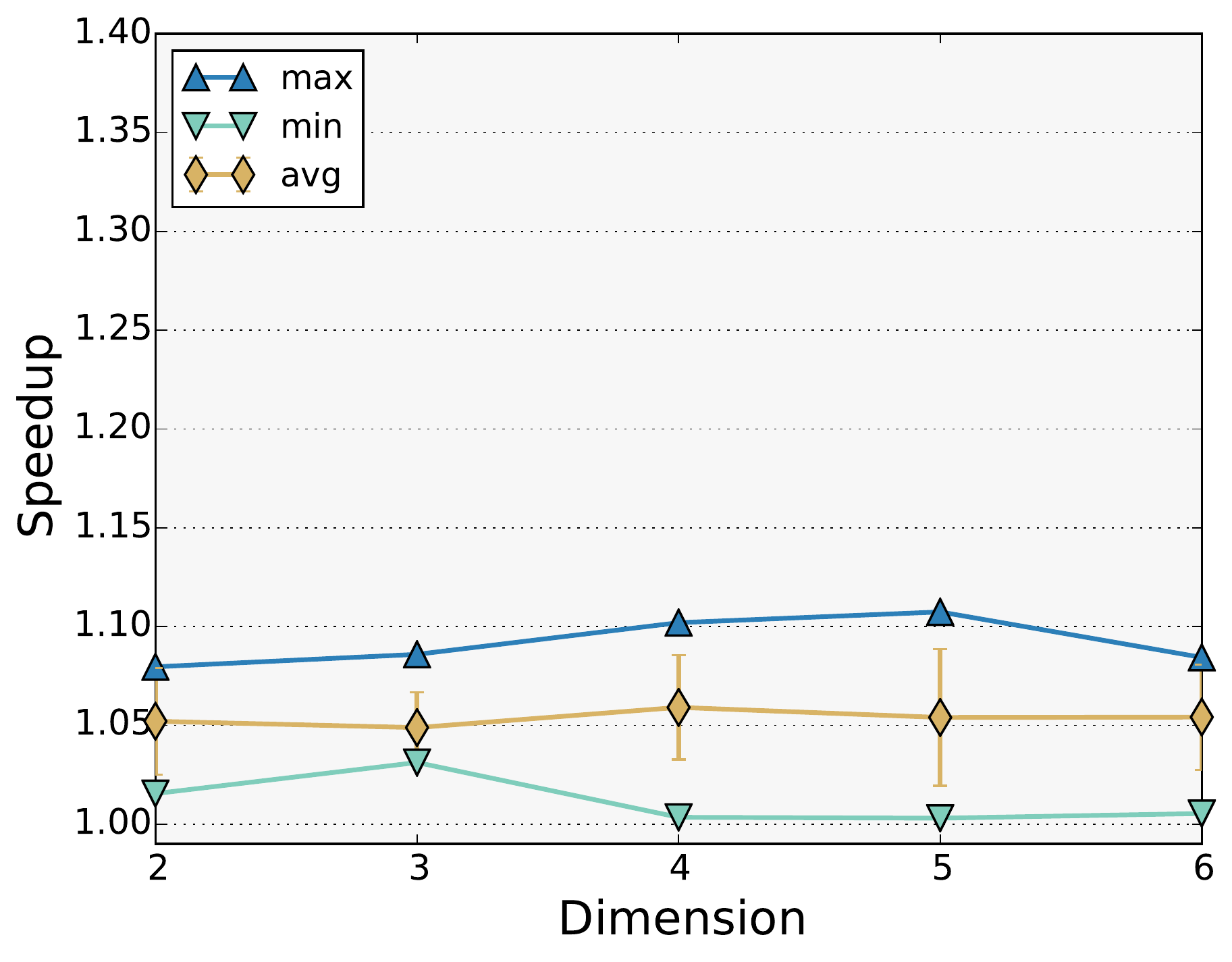}
      \caption{Software prefetching}
      \label{fig:prefetchSpeedup}
   \end{subfigure}\\
   \begin{subfigure}[b]{0.45\textwidth}
   \centering
      \includegraphics[height=4.5cm]{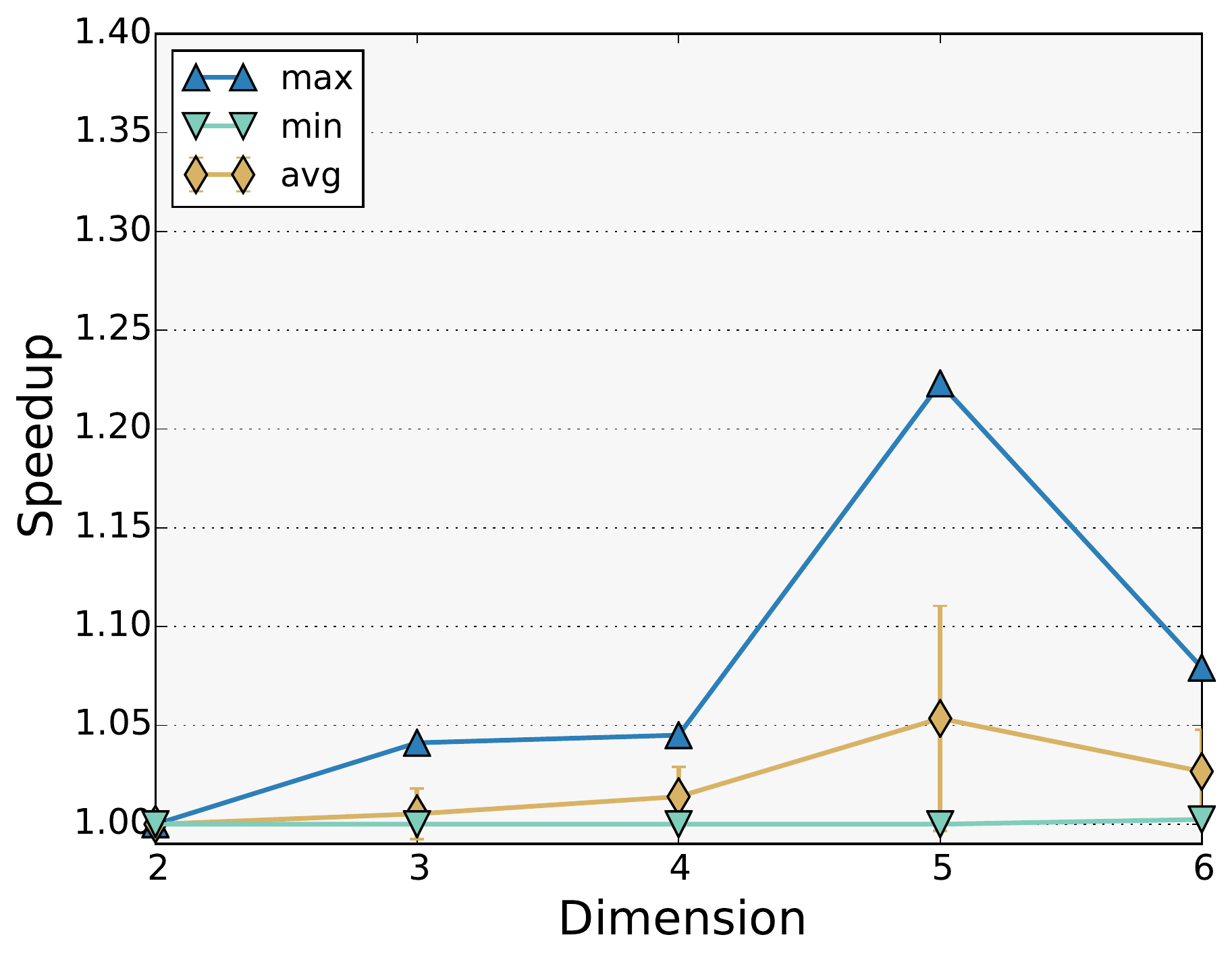}
      \caption{Loop-reordering}
      \label{fig:loopSpeedup}
   \end{subfigure}
   \begin{subfigure}[b]{0.45\textwidth}
   \centering
      \includegraphics[height=4.5cm]{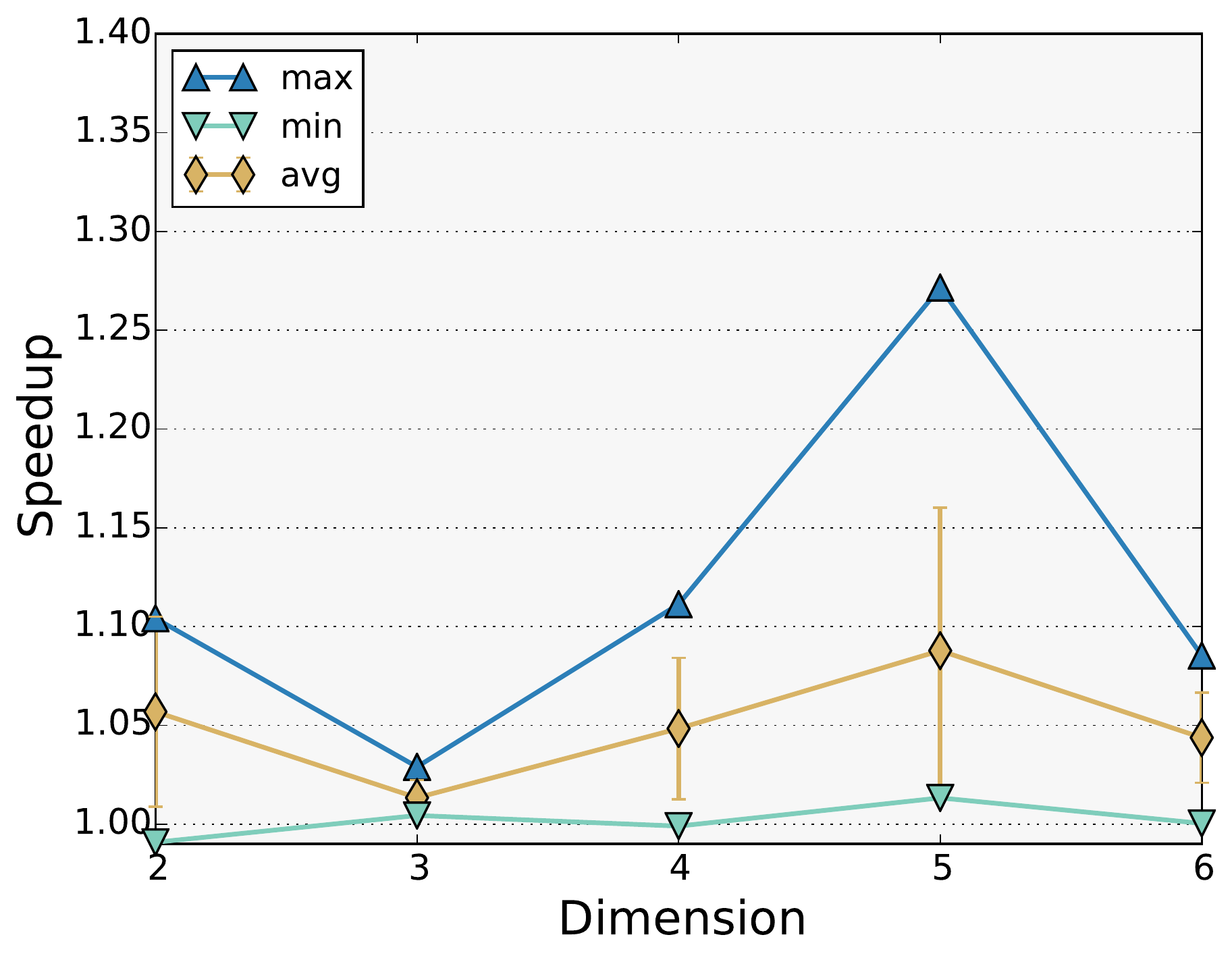}
      \caption{Explicit vectorization}
      \label{fig:vecSpeedup}
   \end{subfigure}
   \caption{Breakdown of the speedups for the \textit{Intel} system.}
   \label{fig:optSpeedup}
\end{figure}

To gain insights on where the performance gains come from, 
in Fig.~\ref{fig:optSpeedup} we report the speedup due to each of TTC's
optimizations separately. 
For each test-case from the benchmark, the speedup is measured as the bandwidth
of the fastest candidate without the particular optimization over the
fastest candidate with the particular
optimization enabled, while all other optimizations are still enabled.\footnote{Note that the remaining parameters of the optimal solutions are
allowed to change (e.g., the fastest solution without software prefetching might
require a blocking of $8\times8$ while a blocking of $16\times16$ is optimal for
solutions with software prefetching enabled).}

Fig.~\ref{fig:blockingSpeedup} presents the speedup that can be gained over a fixed
$8\times8$ blocking. 
The optimal blocking results in up to $35\%$
performance increase and motivates our search in this
search dimension.
Software prefetching (Fig.~\ref{fig:prefetchSpeedup}) also yields a
noticeable speedup of up to $11\%$. 
In contrast to the high speedups gained by loop-reordering shown in Section
\ref{sec:intro}, TTC only exhibits much smaller ones (see
Fig.~\ref{fig:loopSpeedup}).
The reason for this behaviour is twofold: first, we chose a good 
loop order for the reference implementation
(i.e., the loop order for which the output tensor $B$ is accessed in a linear
fashion); second, some of the drawbacks of a suboptimal loop order might be mitigated by
the other optimizations---especially blocking. Even then, an additional search yields speedups of up to $22\%$ over the
reference loop order. 
Despite the fact that transpositions are memory bound, we see an appreciable
speedup by implementing an in-register transpose via AVX intrinsics
(see Fig.~\ref{fig:vecSpeedup}). 
The speedup for vectorization is 
obtained by replacing our explicitly vectorized $8\times8$ micro kernel with a
scalar implementation (i.e., two perfectly nested loops with the
loop-trip-counts being fixed to $8$). While the reference implementation is also vectorized by the
compiler, the compiler fails to find an in-register transpose implementation.

All in all we see that each optimization has a positive effect on the
attained bandwidth; the combination of all these optimizations results in significant
speedups over modern compilers.

\subsection{Reduction of the search space}
We discuss the possibility of lowering the compilation time by reducing the
search space by identifying ``universal'' settings that yield nearly optimal
performance.

%
Intuitively, the optimal prefetch distance (for software prefetching)
should only depend on the memory latency
to the main memory and thus be independent of the actual transposition (see
\cite{lee2012prefetching} for details).
This observation motivates us to seek a ``universal'' prefetch
distance which would reduce TTC's search space.

To evaluate the performance of TTC for a fixed prefetch distance $d$, we introduce
the concept of \textit{efficiency}. Let $C_t$ be the set of all candidates for
the tensor transposition $t$, 
$x$ a particular candidate implementation, and
$d_x$ the prefetch distance used by
candidate $x$; furthermore, let 
$\text{BW}(x)$ be the bandwidth attained by candidate $x$, 
and $\text{BW}^{\text{max}}_t$ the maximum bandwidth among all candidates for
transposition $t$.
Then the \textit{efficiency} $E(d,t)$---which quantifies the loss in performance one would experience if the
prefetch distance were fixed to $d$---is defined as 
\begin{equation} 
   E(d,t) = \max_{\substack{x \in C_t, \\ d_x = d}}\left( \frac{\text{BW}(x)}{\text{BW}^{\text{max}}_t} \right).
\end{equation}
The efficiency is bounded from above and below by 1.0
and 0.0, respectively---with 1.0 being the optimum.

\begin{figure}[t]
   \centering
   \centering
   \includegraphics[width=0.5\textwidth]{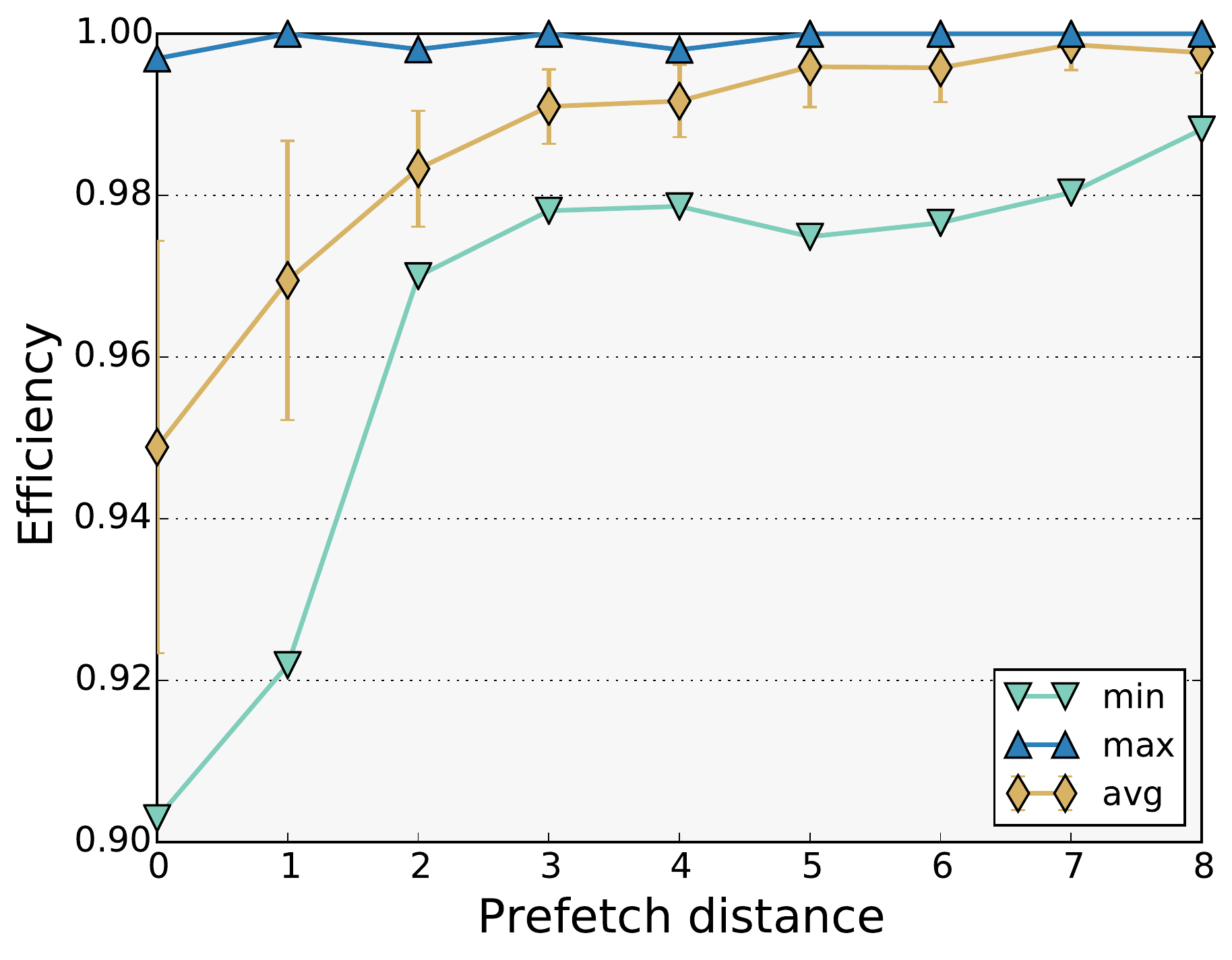}
   \caption{Minimum, average and maximum efficiency for a fixed prefetch distance across the benchmark.}
   \label{fig:slowdownPrefetch}
\end{figure}

Fig.~\ref{fig:slowdownPrefetch} presents the maximum (\includegraphics[height=0.65\baselineskip]{./plots/triangle-up.pdf}), 
minimum (\includegraphics[height=0.65\baselineskip]{./plots/triangle-down.pdf}) and average
(\includegraphics[height=0.8\baselineskip]{./plots/dimond.pdf})
efficiency across all the transpositions of the benchmark as a function of the prefetch distance. 
We notice that (1) there is at least one transposition within the benchmark for
which the influence of software-prefetching is negligible (see the leftmost, blue triangle \includegraphics[height=0.65\baselineskip]{./plots/triangle-up.pdf}),
(2) for each fixed prefetch distance $d$, there is at least one
transposition for which $d$ is suboptimal (see cyan line \includegraphics[height=0.65\baselineskip]{./plots/triangle-down.pdf}),
(3) both the minimum and average efficiency increase with $d$ 
(cyan \includegraphics[height=0.65\baselineskip]{./plots/triangle-down.pdf} and beige \includegraphics[height=0.8\baselineskip]{./plots/dimond.pdf} lines), and (4) once $d$ is ``large enough'',
the efficiency does not improve much.

Quantitatively, a prefetch distance greater or equal to five increases the
average and the minimum efficiency across the benchmark to more than $99\%$ and
roughly $98\%$, respectively. 
Hence, fixing $d$ to any value between 5 and 8 is a good choice for the given
system, effectively reducing the search space by a factor of 9, without
introducing a performance penalty.

\subsection{Quality of Heuristics}
\label{sec:heuristics}

On our Intel system, TTC evaluated roughly $8$ candidates per second across the
whole benchmark (for tensors of size $\SI{200}{\mega\byte}$);
this includes all the necessary steps from code-generation to compilation and
measurement. 
If a solution has to be generated in a short period of time 
(i.e., the search space needs to be pruned efficiently),
the quality of the heuristics to choose a proper loop order and blocking
becomes especially important.

\begin{figure}[t]
   \centering
   \includegraphics[width=0.80\textwidth]{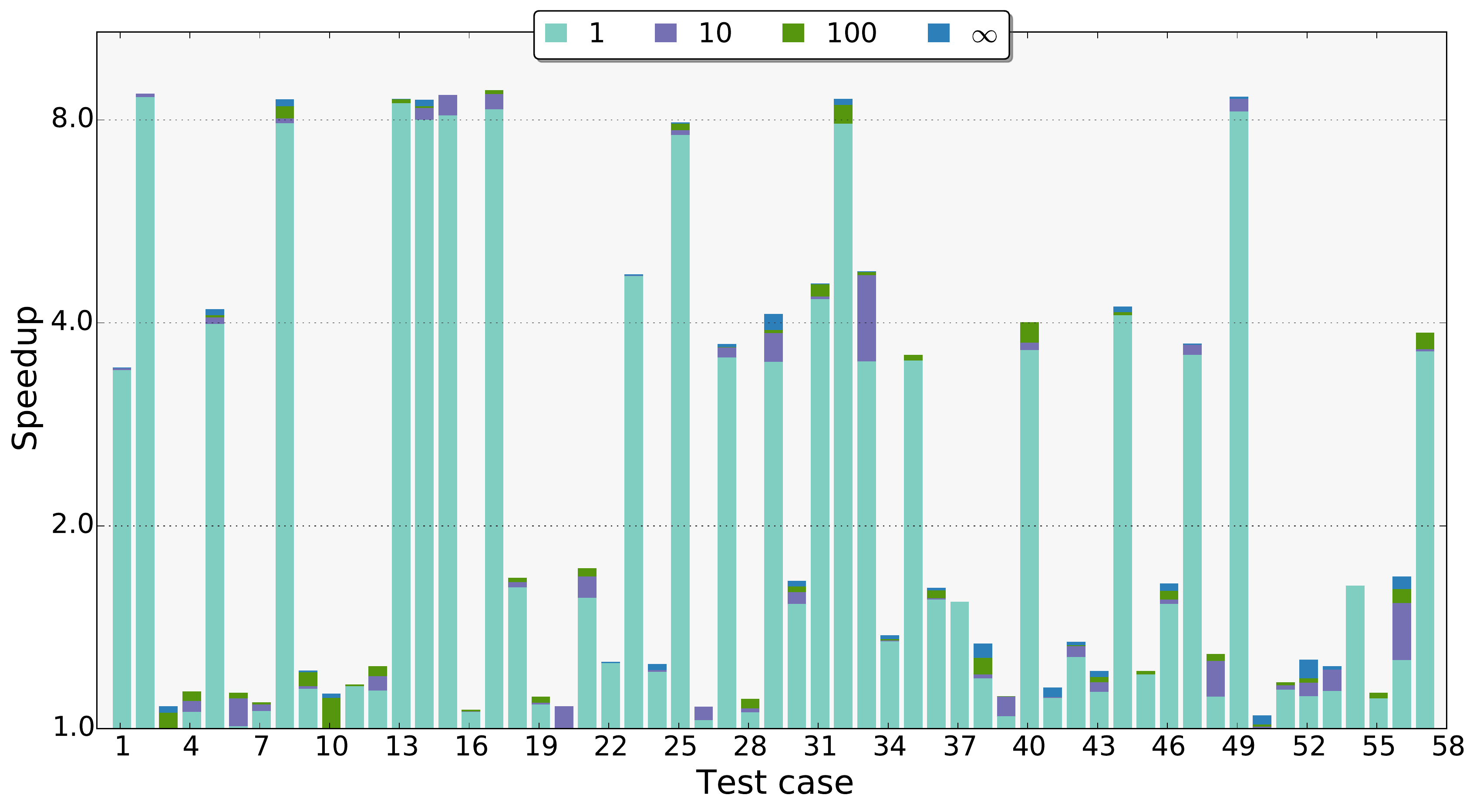}
\caption{Speedup over the reference implementation with the number of
   TTC-generated candidates limited to $1, 10, 100$ and $\infty$ (i.e.,~no
   limit).}
\label{fig:limitedTTC}
\end{figure}

Fig.~\ref{fig:limitedTTC} presents the speedups that TTC achieves if
the user limits the number of generated candidates to $1, 10$ and $100$,
and when no limit is given (i.e.,~$\infty$ reflects
the same results as in Fig.~\ref{fig:benchmarkSpeedupIntel} and
\ref{fig:benchmarkSpeedupAMD}). 
When the number of generated candidates is limited to one,
no search takes place, i.e., the loop order and the blocking for the generated
implementation are determined solely by our heuristics. Even in this extreme
case, TTC still exhibits remarkable speedups over modern compilers.

Specifically, with a search space of $1, 10$ and $100$
candidates, TTC achieves $94.58\%, 97.35\%$ and $99.10\%$ 
of the performance of the unlimited search (averaged across the whole benchmark). 
In other words, one can rely on the heuristics for the most part and resort to search only in scenarios in which 
even the last few bits of performance are critical.
In those cases, we observe that a search space of $100$ candidates
already yields results within $1$\% from those obtained with an exhaustive search.

\section{Conclusion and Future Work}
\label{sec:conclusion}
We presented TTC, a compiler for multidimensional tensor transpositions. 
By deploying various optimization techniques, TTC 
significantly outperforms modern compilers,
and achieves nearly optimal memory bandwidth. 
We investigated the source of the performance gain and illustrated the individual impact of blocking,
software prefetching, explicit vectorization and loop-reordering.
Furthermore, we showed that the heuristics used by TTC efficiently prune the
search space, so that the remaining candidates are easily
ranked via exhaustive search.

For the future, should compilation time become a concern, 
iterative compilation techniques should be considered
\cite{kisuki2000iterative,knijnenburg2002iterative}.
While this current work focused on architectures using the AVX instruction set,
TTC is designed to accommodate other instruction sets; in general, the effort
to port TTC to a new architecture is related to
optimizations such as explicit vectorization and software prefetching.
As a next step, we plan to support the AVX512 instruction set (e.g.,~used by
Intel's upcoming Knights Landing microarchitecture), ARM and IBM Power CPUs as
well as NVIDIA GPUs. Finally, we will be using TTC as a building block for our upcoming Tensor Contraction
Compiler.

\ack{Financial support from the Deutsche Forschungsgemeinschaft (DFG) through grant GSC
111 is gratefully acknowledged.}
\bibliographystyle{ACM-Reference-Format-Journals}
\bibliography{literature}

\end{document}